\documentclass[9pt,sigconf,letterpaper]{acmart}

\usepackage{packages}
\usepackage{layouts}

 \copyrightyear{2022}
 \acmYear{2022}
 \setcopyright{rightsretained}
 \acmConference[IMC '22]{Proceedings of the 22nd ACM Internet Measurement Conference}{October 25--27, 2022}{Nice, France}
 \acmBooktitle{Proceedings of the 22nd ACM Internet Measurement Conference (IMC '22), October 25--27, 2022, Nice, France}
 \acmDOI{10.1145/3517745.3561426}
 \acmISBN{978-1-4503-9259-4/22/10}

\makeatletter
\newcommand\EatSpacesHack{\@bsphack\@esphack}
\makeatother
\iffalse
\newcommand\reviewfix[1]{{\color{red}\sffamily\bfseries [RF:
				#1]}\EatSpacesHack}
			\else
			\newcommand\reviewfix[1]{\EatSpacesHack}
			\fi
\makeatother

\begin{document}

\title[Investigating Apple's New Relay Network]{Towards a Tectonic Traffic Shift?\\Investigating Apple's New Relay Network}

\author{Patrick Sattler}
\orcid{0000-0001-9375-3113}
\affiliation{%
	\institution{Technical University of Munich}
	\country{Germany}
}
\email{sattler@net.in.tum.de}

\author{Juliane Aulbach}
\orcid{0000-0002-0104-5820}
\affiliation{%
	\institution{Technical University of Munich}
	\country{Germany}
}
\email{aulbach@net.in.tum.de}

\author{Johannes Zirngibl}
\orcid{0000-0002-2918-016X}
\affiliation{%
	\institution{Technical University of Munich}
	\country{Germany}
}
\email{zirngibl@net.in.tum.de}

\author{Georg Carle}
\orcid{0000-0002-2347-1839}
\affiliation{%
	\institution{Technical University of Munich}
	\country{Germany}
}
\email{carle@net.in.tum.de}

\ccsdesc[300]{Networks~Network measurement}

\keywords{Relay Networks, Overlay Networks, DNS ECS enumeration}

\begin{abstract}

    Apple recently published its first Beta of the \privaterelay, a privacy protection service with promises resembling the ones of VPNs.
    The architecture consists of two layers (ingress and egress), operated by disjoint providers.
    The service is directly integrated into Apple's operating systems, providing a low entry-level barrier for a large user base.
    It seems to be set up for significant adoption with its relatively moderate entry-level price.

    This paper analyzes the \privaterelay from a network perspective, its effect on the Internet, and future measurement-based research.
    We perform EDNS0 Client Subnet DNS queries to collect ingress relay addresses and find \num{1586} IPv4 addresses.
    Supplementary \ripeatlas DNS measurements reveal \num{1575} IPv6 addresses.
    Knowing these addresses helps to detect clients communicating through the relay network passively.
    According to our scans, ingress addresses grew by \sperc{20} from January through April.
    Moreover, according to our \ripeatlas DNS measurements, \sperc{5.3} of all probes use a resolver that blocks access to \privaterelay. \reviewfix{A.3}

    The analysis of our scans through the relay network verifies Apple's claim of rotating egress addresses.
    Nevertheless, it reveals that ingress and egress relays can be located in the same autonomous system, thus sharing similar routes, potentially allowing traffic correlation.
\end{abstract}

\maketitle

\section{Introduction}

Apple presented \privaterelay~\cite{privaterelay} as a new privacy protection service at its developer conference (WWDC) in June 2021.
The first beta release was included with iOS 15.
This new system seeks to protect its user's privacy by proxying the traffic through an Apple-controlled ingress to an externally controlled egress node.
The service's advertised purpose is to protect the visibility of the user's Internet activity.
It hides the communication partners from passive network observers (\eg \acp{isp}).

The barrier to entry for \privaterelay is much lower compared to competing privacy protection techniques (\eg \acp{vpn}, Proxies, \ac{tor}).
All paying iCloud+ customers can use it on their Apple devices.
Currently, the cheapest iCloud+ plan is \$0.99 per month, which is only a fraction of what, \eg \ac{vpn} operators typically charge.
\privaterelay is integrated into Apple's iOS, iPadOS, and macOS operating systems.
Apple's global smartphone market share is \sperc{25} \cite{phonemarketshare}.
All iCloud+ customers can use the relay network by flipping a switch in their device settings.
Apple also announced that it would turn on the service by default after it left the beta testing phase.
We expect to see a significant usage increase when that happens.
The ease-of-use aspect, the hidden metadata, and a potential broad adoption scenario are also alarming to the actors in the networking community (see \cite{telegraph-isps-whining}).

This paper analyzes the goals, architecture, and behavior of \privaterelay to provide fellow researchers and network operators insight into the system and what they can expect when the service adoption gains traction.
\privaterelay influences the operation of passive network analysis, and due to its workings, it can also impact \acp{ids}. \reviewfix{A.2}
Its structure also allows the participating \acp{cdn}, which host the egress nodes, to use their involvement in this system to provide content hosted with them faster than 3rd-party content.
The focus of this paper is the network point of view, and where applicable, we evaluate the effect of our findings on the system's privacy and security.

\vspace{0.2em}
\noindent\textbf{This paper provides the following contributions:}

\noindent
\first We analyze the \privaterelay ingress layer and collect \num{1586} IPv4 and \num{1575} IPv6 ingress relay IP addresses through active \ac{dns} scans.
The ingress addresses can be used to identify relay traffic as a passive network observer.
Researcher can use our published results in their analysis.

\noindent
\second We evaluate the egress addresses and show their geographical and topological bias towards the US.
The egress addresses provide us with a better understanding of the service's deployment status.

\noindent
\third We perform active scans through the relay network and find that the egress relay not only rotates its address but also does this for every new client connection.
Services similar to \privaterelay (\eg \ac{vpn}, \ac{tor}) do not exhibit such a behavior.
Therefore, \acp{ids} need to consider this new type of connection pattern.
The scans also reveal that ingress and egress addresses can be located behind the same last-hop router, enabling the network operator to perform correlation attacks similar to the ones known for \ac{tor}~\cite{murdoch-tor-traffic, feamster-locating-tor, overlier-locating-servers}.

\noindent
\fourth In \Cref{sec:discussion} we discuss our results considering a broader usage in the future. We consider passive network analyses, the topological location of relay nodes, and the impact on network defense systems. \reviewfix{A.1} \reviewfix{A.2}

\vspace{0.2em}

We publish data used throughout this work as a research data archive~\cite{datatum} and provide current and future measurement results at: \textit{\url{https://relay-networks.github.io}}.

\vspace{0.2em}
\noindent\textbf{Outline:} We introduce \privaterelay in detail in \Cref{sec:background} and describe our measurement setup and scans in \Cref{sec:scans}.
We analyze ingress relays, egress relays, and their interplay in \Cref{sec:results}.
In \Cref{sec:related} we list related work, and \Cref{sec:discussion} concludes the paper with a discussion on our findings and the design of \privaterelay.

\section{\privaterelay}
\label{sec:background}

\begin{figure}
    \includegraphics{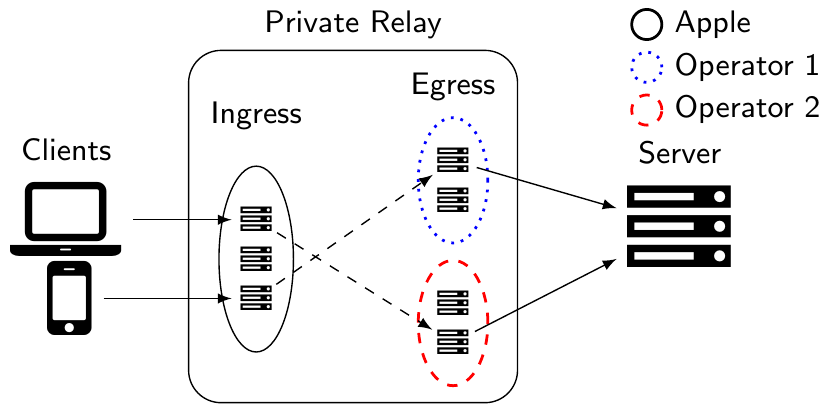}
    \caption{\privaterelay Architecture. Ingress nodes are operated by Apple itself, while egress nodes are operated by different entities. Only the ingress can identify the client whereas only the egress knows the targeted service.}
    \label{fig:private-relay-arch}
\end{figure}

\privaterelay is a new service provided by Apple~\cite{privaterelay} in order to protect the user's privacy.
It aims to protect unencrypted \acs{http} traffic, \ac{dns} queries, and connections initiated by Apple's web browser Safari.
Moreover, no network operating party can observe the client and server addresses directly.
The architecture of \privaterelay (see \Cref{fig:private-relay-arch}) is built as a two-layer relay structure as lined out in its whitepaper~\cite{privaterelay}.
Clients, \ie iOS or MacOS devices, connect to an ingress relay for authentication and location assignment.
The clients use this information to initiate a proxy connection to the egress relay through the ingress.
The latter then initiates the connection to the actual target host.
Apple \emph{operates} the ingress layer relays, while \ac{cdn} providers Akamai, Cloudflare, and Fastly operate the egress relays (\cf \Cref{sec:egress}).
Note that while Apple states it operates the ingress layer nodes, \Cref{sec:ingress} shows that these are not necessarily located in a network Apple operates. \reviewfix{D.2}

This architecture enables several advantages compared to traditional proxy or \ac{vpn} services, \eg the egress layer can initiate the connection and use additional latency-reducing techniques (\eg using TCP fast open).
Cloudflare, an egress relay operator, claims to use Argo~\cite{cf-icpr}, its \emph{virtual backbone} which analyzes and optimizes routing decisions~\cite{argo} and improves connection performance.
We assume the other \ac{cdn} egress operator have similar measures in place.
These measures might be enough to equalize any latency drawbacks due to the two-hop relay system.

Proxying traffic through the ingress relay protects the client's IP address from the egress relay, the destination server, and the network operating entities on the path between the relay to the destination.
Conversely, the ingress relay cannot decrypt traffic to the egress and is, therefore, unable to observe the destination address or any other information.
Network operators between client and ingress are also unable to observe the actual destination address.
This layered structure has similarities with anonymization networks such as \ac{tor}.
\ac{tor} uses at least three layers, and volunteers independently operate its relays.
In this paper, we will apply analysis approaches similar to the ones used for \ac{tor} localization and evaluation \cite{nasr2018tor,sun-raptor,johnson-tor-routing,murdoch-tor-traffic}.

\privaterelay uses QUIC for its connection to the ingress relay.
The tunnel to the egress uses a secure proxying protocol using HTTP/3~\cite{kuehlewind-masque, masque_draft} proposed by the \ac{masque} IETF working group~\cite{masque_wg}.
HTTP/3 uses QUIC as the transport protocol providing a secure connection with the possibility to combine multiple connections within a single proxy connection.
The service uses the fallback to HTTP/2 and TLS 1.3 over \ac{tcp} when the QUIC connection fails.
Clients resolve \texttt{mask.icloud.com} to obtain the ingress relay's addresses for the QUIC connection.
During the \ac{tcp} fallback, clients resolve \texttt{mask-h2.icloud.com}.
The whitepaper explicitly states the possibility of blocking \privaterelay by not resolving \ac{dns} requests for the service's domain names.

To authenticate servers, \privaterelay relies on \emph{raw public keys} (see RFC~7250~\cite{rfc7250}) instead of the usual certificate authentication within its TLS handshake.
Deployed key pinning prevents interception using TLS proxies, so an in-depth analysis of the protocol is infeasible.
Additional measures for fraud prevention are in place, \eg a limited number of issued tokens to access the service per user and day.

\privaterelay is unavailable in some countries where local laws do not permit Apple to operate it (\eg China, Belarus, and Saudi Arabia \cite{prnotavailablechina}). \reviewfix{D.3}
The current system claims to not use any network block circumvention mechanisms as \ac{tor} and \ac{vpn} services often do.
\privaterelay can easily be blocked through its domain names.
Differences compared to other tunneling services are that \privaterelay does not apply to all traffic.
Moreover, it uses the \ac{masque} proxying technique. \reviewfix{D.1}
Currently, proxying UDP traffic is not supported by \ac{masque}, but the \ac{masque} working group is working on a new draft~\cite{draft-ietf-masque-connect-udp}.

\section{Measurement and Analysis Setup}
\label{sec:scans}
We aim to shed light on the network-level implementation of the \privaterelay and its inner workings.
The ingress and egress relays are the visible points of the system from an external perspective.
Collecting and understanding the properties of the relays is essential to understand the systems deployment and which prefixes and addresses are relevant for network analysis. \reviewfix{A.4}
Thus, we analyze ingress and egress relays regarding their topological distribution from a network perspective (\ie used addresses, prefixes, and operators).
For all active measurements, we apply the ethical measures described in \Cref{sec:ethics}.

\textbf{Relay IP Addresses:}
Apple does not publish used ingress IP addresses, unlike egress addresses.
As ingress addresses are the relay network's entry point, it is crucial to obtain them.
They can detect the presence and prevalence of relay network traffic.
We rely on \ac{dns} queries for the domains used by \privaterelay to obtain the ingress IP addresses.
The \privaterelay name servers are operated by AWS Route 53 and have the \ac{ecs} extension enabled.
\ac{ecs} allows a resolver to attach the client's subnet to a \ac{dns} query \cite{rfc7871}.
The authoritative name server can use this query information to provide a subnet-based (\ie geolocation-based) response.

\citeauthor{streibelt-ecs}~\cite{streibelt-ecs} and \citeauthor{calder2013ecs}~\cite{calder2013ecs} first described \ac{ecs} enumeration approaches. \reviewfix{A.5}
Our scan iterates over the IPv4 address space and sends A record queries with /24 subnets in its \ac{ecs} extension to the authoritative name server.
This ECS-based approach does not work for IPv6.
The name server always returns an \ac{ecs} scope indicating that the response is valid for the entire IPv6 address space. \reviewfix{E.1}

To cover IPv6 (\emph{AAAA} queries) and to verify the results obtained by the ECS-based scans, we use \ripeatlas \ac{dns} measurements, offering globally distributed probes and thus a geographically distributed view.
\ripeatlas \emph{A} queries are used to validate our ECS-based \ac{dns} scans.
Additionally, we use the \emph{A} and \emph{AAAA} measurements to track the service's availability.
We use the \ripeatlas \ac{dns} resolutions to gain insight into the number of probes failing to resolve the service domain names.
Even though \ripeatlas probes are not distributed equally~\cite{bajpai2015atlas}, they are located in \num{3326} different \acp{as} and \num{168} countries and, therefore, can provide us with a distributed view. \reviewfix{C.1}
Tracking the availability is an important future task as the service can be easily blocked through \ac{dns}.
Several ISPs voiced concerns~\cite{telegraph-isps-whining} over the service, and some might start blocking it in their network.

In contrast to ingress relays, Apple publishes egress relay IP addresses for geolocation and allow-listing \cite{egress-nodes}.
We assume this list to be complete and use it in the following.

\textbf{Measurements using \privaterelay:}
We perform several scans to improve our understanding of the system outside the published information.
As explained in \Cref{sec:background}, we cannot examine the communication itself due to the pinned public key.
In fact, testing standard QUIC handshakes using the QScanner published by \citeauthor{zirngibl2021over9000} \cite{zirngibl2021over9000} or a current curl version does not even trigger a response by ingress nodes, neither a QUIC initial nor an error.
The connection attempt times out.
Interestingly, a version negotiation from ingress nodes can be triggered using the latest ZMap module from \citeauthor{zirngibl2021over9000} \cite{zirngibl2021over9000} to identify QUIC support.
The response indicates support for QUICv1 alongside drafts 29 to 27.
These response properties verify that nodes support standardized QUIC, but due to its peculiarities, unintended handshakes are prevented.

Instead, we perform long-running measurements on a MacBook Pro laptop with \privaterelay enabled to understand how often the egress operator and IP address change.
The service emphasizes privacy as the primary goal and announces to rotate the egress IP address regularly, a feature unique among similar services (\eg \ac{vpn}, \ac{tor}).
The address rotation hinders IP address-based tracking significantly more than without it. \reviewfix{A.6}
For this, we set up a MacOS laptop running the latest OS version as a client for the \privaterelay and deployed a simple web server to observe the egress relay's connection attempt.
The scan performs two requests: \first It instructs \emph{Safari} to open the URL to our web server; and \second we use curl to fetch \texttt{http://ipecho.net/plain}, a service that mirrors the requester's IP address.
We directly log the requester's IP address on our web server and extract the IP address from the response of \texttt{http://ipecho.net/plain}.\reviewfix{ShadowPC}
Therefore, we can observe and correlate ingress \emph{and} egress relay operator and evaluate how parallel connections behave.

We implement this scan in two versions.
\noindent
\first An open scan, where required DNS queries are sent to a local recursive resolver to initiate the \privaterelay connection.
Thus, this scan uses IP addresses received live from authoritative name servers.
\noindent
\second We perform scans forcing the client to use specific \ac{dns} configurations.
Therefore, we deploy and use a local unbound resolver to steer the service's \ac{dns} resolution.
A custom configured local zone for the required domain names can direct the service to a selected ingress relay.
Such a forced ingress selection allows us to test the relay's behavior when using different IP addresses from our \ac{ecs} scan results as ingress nodes.

\section{Analysis of Scanning Results}
\label{sec:results}
In the following section, we analyze the topological distribution and inner workings of the \privaterelay.

\subsection{Uncovering Ingress Relays}
\label{sec:ingress}
We use the \ac{ecs} scans to uncover globally distributed ingress addresses from a single vantage point and verify its results in the second part of this section by using \ripeatlas measurements.

\textbf{\ac{ecs} \ac{dns} scans:}
\Cref{tbl:ingress-shares} gives an overview of the number of seen ingress IP addresses per \ac{as} from four scans between January and April 2022. %
Ingress addresses are located at Apple (AS714) or Akamai (AS36183) and within \num{123} routed BGP prefixes.
AS36183 appears to be only recently active in BGP and is related to \privaterelay.
We use \akamaipr to denote AS36183.
The increase of ingress addresses is solely attributable to \akamaipr.
Especially, the fallback TCP relays were initially served by Apple, and only after the deployment of relays at \akamaipr the fallback relays could catch up with the QUIC relays.

\begin{table}
    \footnotesize
    \centering
    \caption{The \acp{as} of ingress relays and their proportional distribution. Only Apple and the recently occurring \akamaipr AS (AS36183) is visible. In January the fallback scan is absent.}
    \label{tbl:ingress-shares}
    \begin{tabular}{lS[table-format=4.0]S[table-format=2.1, table-space-text-post = \si{\percent}]S[table-format=4.0]S[table-format=2.1, table-space-text-post = \si{\percent}]S[table-format=4.0]S[table-format=2.1, table-space-text-post = \si{\percent}]S[table-format=4.0]S[table-format=2.1, table-space-text-post = \si{\percent}]}
        \toprule
        & \multicolumn{4}{c}{Default} & \multicolumn{4}{c}{Fallback} \\
        \cmidrule(lr){2-5} \cmidrule(lr){6-9}
        & \multicolumn{2}{c}{Apple} & \multicolumn{2}{c}{Akamai} & \multicolumn{2}{c}{Apple} & \multicolumn{2}{c}{Akamai} \\
        \cmidrule(lr){2-3} \cmidrule(lr){4-5} \cmidrule(lr){6-7} \cmidrule(lr){8-9}
        Jan & 365 & 30.6 \si{\percent} &  823 & 69.4 \si{\percent} & {-} &   {-} &  {-} &  {-} \\
        Feb & 355 & 29.5 \si{\percent} &  845 & 70.5 \si{\percent} & 356 & 100.0 \si{\percent} &  {-} &  {-} \\
        Mar & 347 & 26.9 \si{\percent} &  945 & 73.1 \si{\percent} & 334 &  93.0 \si{\percent} &   25 &  7.0 \si{\percent} \\
        Apr & 349 & 22.0 \si{\percent} & 1237 & 78.0 \si{\percent} & 336 &  24.0 \si{\percent} & 1062 & 76.0 \si{\percent} \\
        \bottomrule
    \end{tabular}
\end{table}

\begin{table}
    \setlength{\tabcolsep}{3.8pt}
    \centering
    \footnotesize
    \caption{Number of client \acp{as} served by each ingress relay \ac{as} for the scan in April. The \ac{as} population is sourced from the APNIC AS pop data \cite{aspop}.}
    \label{tbl:operator_shares}
\begin{threeparttable}
    \begin{tabular}{lS[table-format=4.0, table-space-text-post = \si{\mn}]S[table-format=4.0]S[table-format=2.1, table-space-text-post = \si{\mn}]S[table-format=3.0, table-space-text-post = \si{\percent}]}
    \toprule
    {AS} & {AS Pop} & {\acp{as}} & {/24 Subnets} \\
    \midrule
    Akamai$_{PR}$ & 994 \si{\mn} & 34627 & 1.1 \si{\mn} \\
    Apple & 105 \si{\mn} & 20807 & 0.2 \si{\mn} \\
    Both\tnote{1} & 2373 \si{\mn} & 17301 & 10.6 \si{\mn} \\
    \bottomrule
    \end{tabular}
\begin{tablenotes}
    \item[1] Apple's subnet share is \sperc{76}
\end{tablenotes}
\end{threeparttable}
\end{table}

In April, our scan uncovered \num{1586} ingress IP addresses in responses with up to eight different records.
\akamaipr locates more than \sperc{75} of all relays.
Therefore, the question arises: How reliant is \privaterelay from \akamaipr, \ie who serves its clients?
The \ac{ecs} scan collects this information as the sent extension data represents the client's subnet.
The name server always uses the subnet provided in the query, and all response records are in the same \ac{as}.
As we send /24 client subnets, we obtain a /24 prefix granularity.
\Cref{tbl:operator_shares} summarizes the collected data on \ac{as} and subnet levels.

Although Apple only provides \sperc{25} of all ingress IP addresses, it serves \sperc{69} of all subnets.
We use the APNIC AS population dataset \cite{aspop} to understand how many potential users are served by the two operators.
As the population dataset has only an \ac{as} granularity, \acp{as} served by both cannot be attributed.
These \acp{as} contain the largest share of subnets and users.
The relays at \akamaipr exclusively cover \sk{34.6} \acp{as} with \sm{994} users.
In comparison, relays at Apple only cover \sm{105} users in \sk{20.8} \acp{as}.
This breakdown shows that the deployment in the Apple \ac{as} differs from the one in \akamaipr.
If network traffic is evaluated on a service level, it is crucial to consider this split-world for \privaterelay.
Our ingress address dataset helps to attribute IP addresses and aggregate service level data.

\privaterelay evolved during our four-month observation period:
The number of QUIC-enabled relays increased by \sperc{34}, and the number of relays for the TCP fallback increased from only \num{356} addresses to \num{1398} (+\sperc{293}) in April.
Such a development is expected for a service in its beta phase.
It shows the importance of continuous measurements to understand its impact.
We will perform regular scans in the future and publish the collected ingress addresses.

\textbf{\ac{ecs} Scan Validation:}
The ECS-based ingress scan uses the \ac{dns} extension's properties to get a distributed view of the service.
We perform these scans from a single vantage point.
Therefore, they are susceptible to anycast-based behavior differences.
To validate our results, we use the \ripeatlas measurement platform, which allows us to perform \ac{dns} measurements on more than \sk{10} available probes.
The \ripeatlas probe location bias towards North America and Europe is similar to the egress subnet locations (\cf \Cref{sec:egress}).
Therefore, we acknowledge the uneven distribution of \ripeatlas but argue that it also roughly represents the current service deployment of \privaterelay.

In total, our \ripeatlas measurements in April report \num{1382} distinct ingress IPv4 addresses, \ie 200 fewer than the \ac{ecs} scan at a similar time.
All but one address from the \ripeatlas measurement are also visible in our \ac{ecs} scan. \reviewfix{E.2}
The single missing address can be attributed to the time difference between the two scans.
While the \ripeatlas scan only takes minutes, our \ac{ecs} scan takes up to 40 hours due to the strict rate limiting.
We can find this single missing address during the following verification \ac{ecs} scans.
We conclude that our \ac{ecs} scan can uncover not only all addresses seen by the \ripeatlas measurement but also 200 additional ones.

\textbf{IPv6 Ingress Addresses:}
We also perform four \emph{AAAA} \ac{dns} scans with \ripeatlas as our \ac{ecs} scan only supports IPv4 with A type records.
We use \emph{AAAA} measurements targeting the local resolver and  an authoritative name server.
According to our measurement towards \texttt{whoami.akamai.net}, which returns the resolver's requester IP address, more than half of all probes use one of the following resolvers: Google's public resolver \cite{google-pub-dns}, Cloudflare's public resolver \cite{cloudflare-dns}, Quad9 \cite{quad9}, or OpenDNS \cite{opendns}.
Although, the geographical bias of \ripeatlas probes and the concentrated public resolver usage limits the visibility of this scan, resolvers are visible in \sk{1.8} different \acp{as}. \reviewfix{E.3}

In total, we find \num{1575} IPv6 addresses in the same two \acp{as}.
The size of the discovered IPv6 address set through \ripeatlas is larger than the IPv4 one, possibly due to the larger address space.
The \ac{as} share of addresses is similar to what we find in our \ac{ecs} scans: \num{346} relays are within Apple's \ac{as} and \num{1229} relays are provided by the \akamaipr \ac{as}.
The \ac{dns} requests directed to an authoritative name server do not expose significantly more or other addresses as the resolver scan does.
This measurement provides us basic view into the IPv6 side of the service.

\textbf{State of Service Blocking:}
We analyze the \ripeatlas \ac{dns} errors to check where the \ac{dns} resolution fails and where the service might be blocked.
\sperc{10} of all requested probes fail with a request timeout.
An additional measurement towards another domain showed similar timeout shares.
Therefore, we do not account these as service blocking attempts.
Nevertheless, \sperc{7} of the probes fail to resolve the domain name but receive a \ac{dns} response by their resolver. %
\sperc{72} of these probe's response codes are \emph{NXDOMAIN}, \sperc{13} \emph{NOERROR}, and \sperc{5} \emph{REFUSED}.
The remaining ones report either \emph{SERVFAIL} or \emph{FORMERR}.
Responses with \emph{NXDOMAIN} or \emph{NOERROR} with no data are responses where the resolver claims to have completed the resolution to the authoritative name server and to have returned its result.
We know that the authoritative name server does not respond with any of the results above.
Therefore, we attribute these response codes as intentional blocking of the \privaterelay domain names. \reviewfix{E.4}
In one occurrence we observe a \ac{dns} hijack hinting at the use of \url{nextdns.io}, a \ac{dns} resolver claiming to protect from different Internet threats.
The instances returning \emph{REFUSED} might also be caused by erroneous \ac{dns} setups but in our case we verified the functioning of the resolver using a second unrelated domain.
Therefore, we find a total of \num{645} (\sperc{5.5}) probes without access to the service due to \ac{dns} blocking.

\subsection{Egress Nodes}
\label{sec:egress}

\begin{table}
	\setlength{\tabcolsep}{3.8pt}
	\begin{threeparttable}
		\footnotesize
		\centering
		\caption{Comparison of egress subnets for the operating \acp{as}. Number of possible IPv6 addresses is left out, since every subnet mask has length 64.}
		\label{tbl:egress_asn}
		\begin{tabular}{lrrrrrr}
			\toprule
			& \multicolumn{3}{c}{IPv4} & \multicolumn{2}{c}{IPv6} \\
			\cmidrule(lr){2-4} \cmidrule(lr){5-6}
			& Subnets & \acs{bgp} Pfxs & IP Addr. & Subnets & \acs{bgp} Pfxs & \acsp{cc}\\
			\midrule
			Akamai$_{PR}$\tnote{1}      & \num{9890} & \num{301} & \num{57589} & \num{142826} & \num{1172} & \num{236} \\
			Akamai$_{Est}$\tnote{2}     & \num{1602} & \num{  1} & \num{ 5100} & \num{ 23495} & \num{   1} & \num{ 24} \\
			Cloudflare\tnote{3}         &\num{18218} & \num{112} & \num{18218} & \num{ 26988} & \num{   2} & \num{248} \\
			Fastly\tnote{4}             & \num{8530} & \num{ 81} & \num{17060} & \num{  8530} & \num{  81} & \num{236} \\
			\bottomrule
		\end{tabular}
		\begin{tablenotes}
			\item[1] AS36183; \item[2] AS20940; \item[3] AS13335; \item[4] AS54113
		\end{tablenotes}
	\end{threeparttable}
\end{table}

\begin{figure*}
	\begin{subfigure}{.33\textwidth}
		\centering
		\includegraphics[width=.9\linewidth]{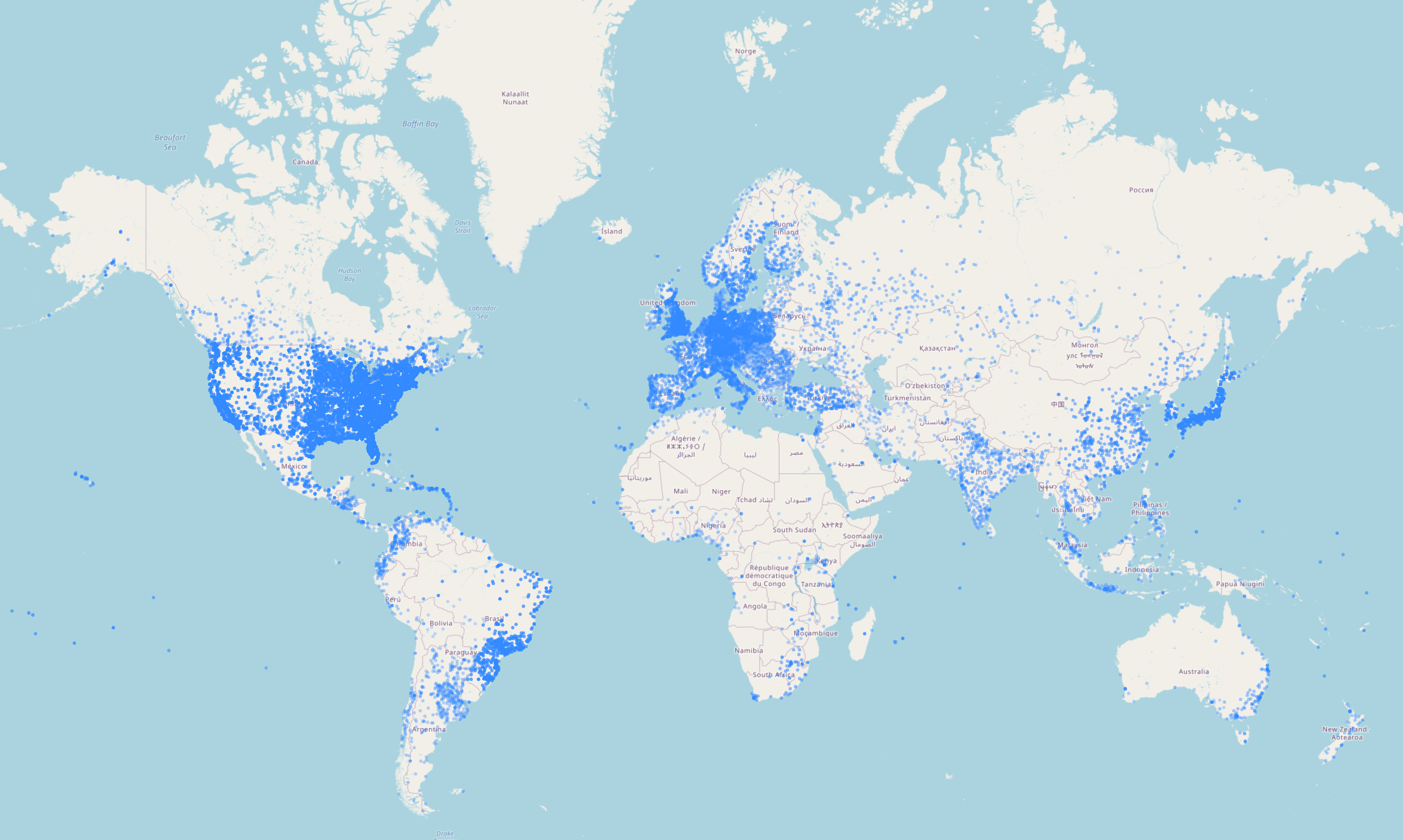}
		\caption{\akamaipr and \akamaiest}
		\label{fig:sub1}
	\end{subfigure}%
	\begin{subfigure}{.33\textwidth}
		\centering
		\includegraphics[width=.9\linewidth]{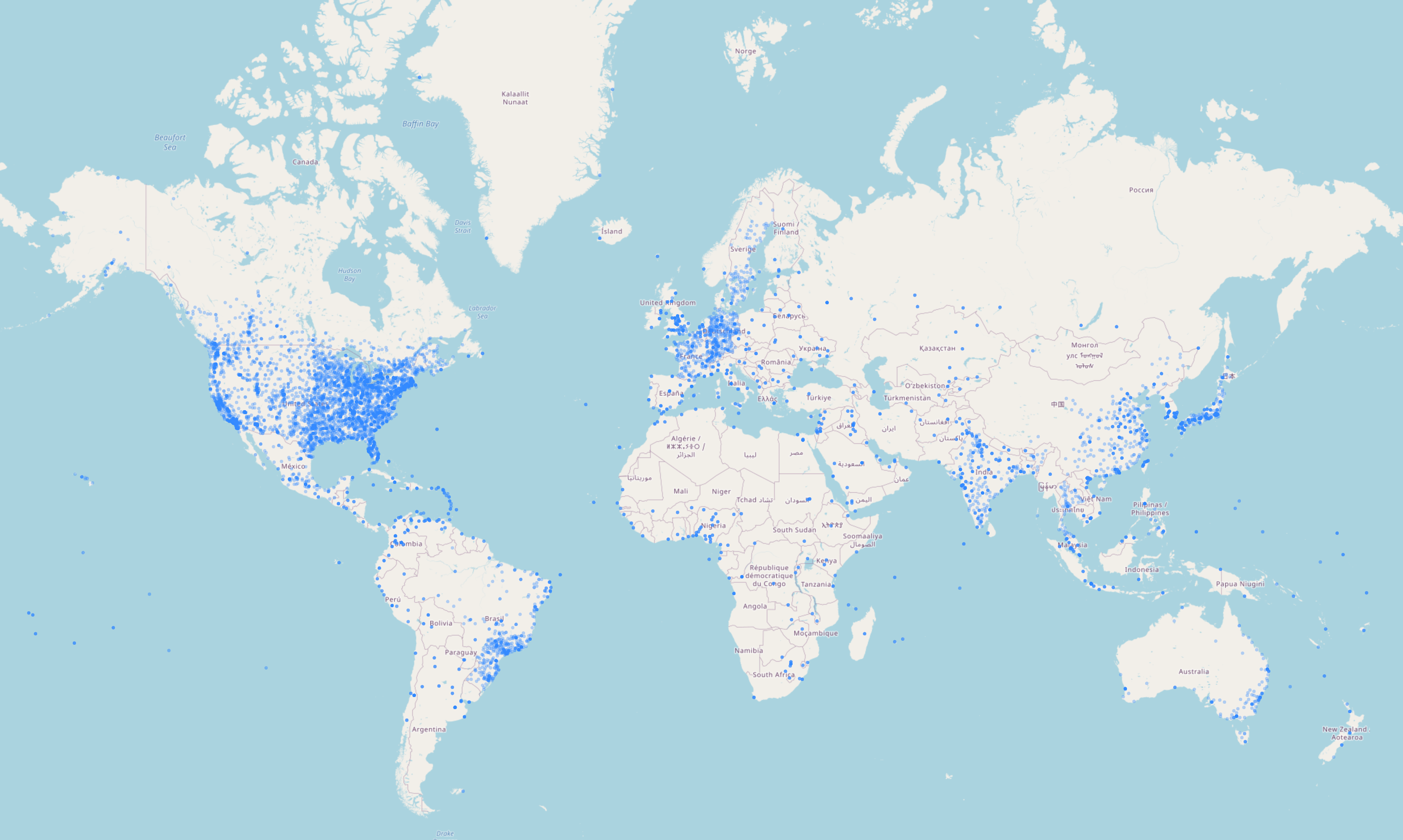}
		\caption{Cloudflare}
		\label{fig:sub2}
	\end{subfigure}
	\begin{subfigure}{.33\textwidth}
		\centering
		\includegraphics[width=.9\linewidth]{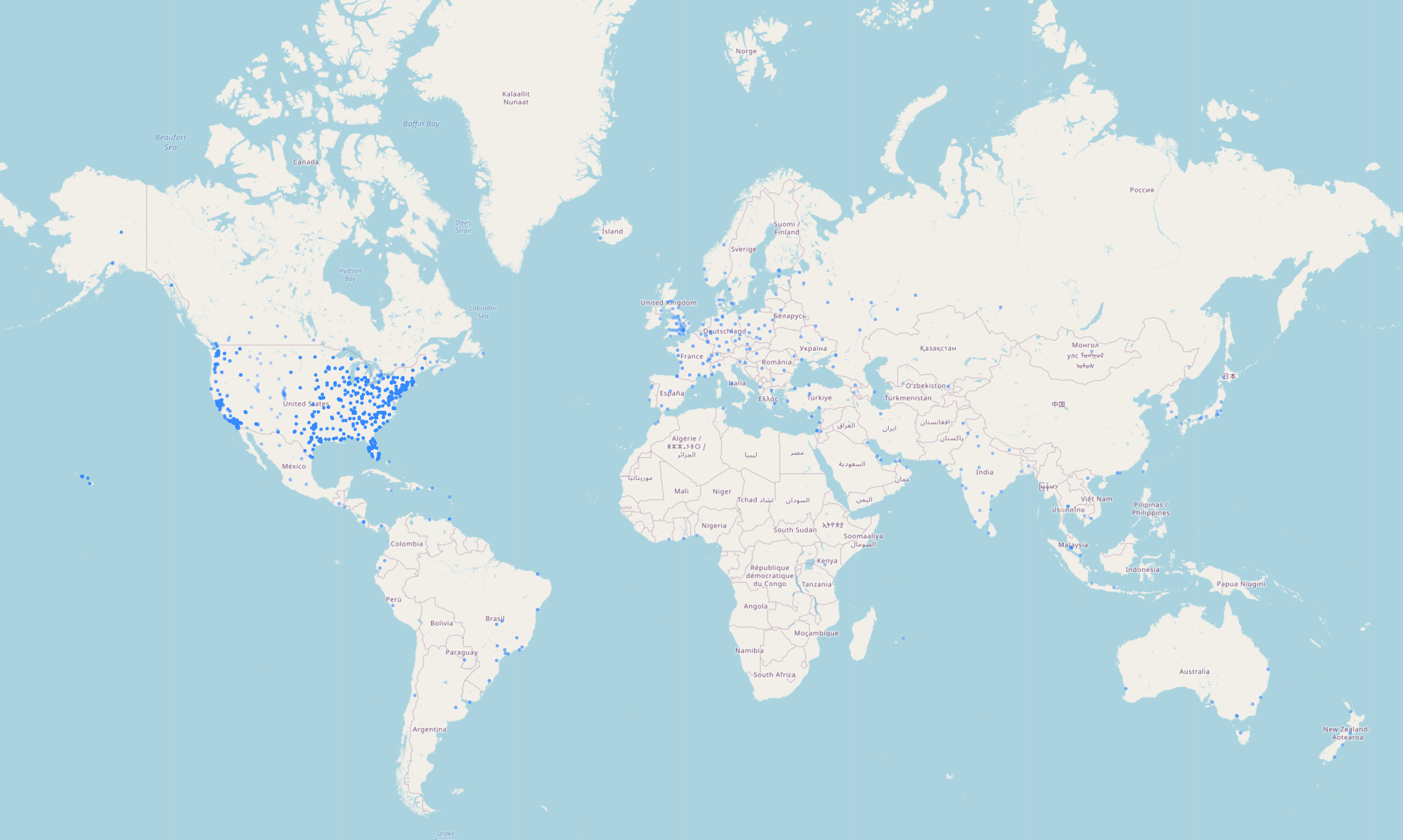}
		\caption{Fastly}
		\label{fig:sub3}
	\end{subfigure}
	\caption{Geolocation of egress subnets per providing \ac{as}\protect\footnotemark.}
	\label{fig:node_dist}
\end{figure*}
The egress node is the second hop of the \privaterelay infrastructure.
Depending on the selected option, the node either maintains the user's region by leveraging geohash information or only preserves the country and time zone.
Apple provides a list of the egress nodes \cite{egress-nodes}
that currently (2022-05-11) contains \sk{238} subnets, each mapped to a represented \ac{cc}, region, and city.
For \sperc{1.6} of the subnets, the city is missing.
We assume those subnets are used if a user does not want to maintain the region, leaving it blank for our analysis.
Compared to Jan. 2022, the number of subnets grew by \SI{15}{\percent} with little churn.

\textbf{\ac{as} Distribution:}
As depicted in \Cref{tbl:egress_asn} the subnets are all in the \acp{as} of Akamai, Cloudflare, and Fastly, with Akamai being represented by two different \acp{as}, namely \akamaipr (AS36183) and AS20940.
We refer to the latter as \akamaiest.
Interestingly, \akamaipr also hosts ingress nodes, as shown in \Cref{sec:ingress}, thus combining both layers of the \privaterelay within the same \ac{as}.
While Cloudflare offers the most IPv4 subnets, Akamai provides more possible IP addresses.
Regarding IPv6, Akamai is offering the most subnets.
All listed IPv6 subnets by Apple have a \num{64} bit subnet mask.
Hence no number of addresses is explicitly given. %
For the \num{9890} subnets of \akamaipr, we see \num{301} different routed BGP prefixes, whereas \akamaiest routes all \num{1602} subnets over the same \acs{bgp} prefix.
This single IPv4 BGP prefix contains subnets covering 18 countries distributed over North and South America and Europe.
Even though the egress relays could theoretically all be placed at the Apple-provided location, it does not seem likely as to get low latency relays have to be located in a topologically convenient place.
Akamai publishes a list of countries with points of presence.
We compared this list to the countries in the egress list and found several small countries (\eg Saint Kitts and Nevis) with a representing IP address from Akamai but without any point of presence. \reviewfix{B.3}
This analysis shows that the published location information does not necessarily represent the egress node's actual location but is used to represent the client's assumed location.

We also used the MaxMind GeoLite2 geolocation database to obtain the location for the egress IP addresses and found that they adapted the Apple egress mapping for most subnets.
Among others, MaxMind advertises its GeoIP databases to be used for content customization and advertising.
Therefore, their goal is to represent the user's location, not the relay node's actual location, and thus these databases cannot be used to determine the relay node's location.

\textbf{Geo Distribution:}
\Cref{fig:node_dist} shows the distribution of egress subnets over the globe.
Cloudflare provides egress relays in \num{248} and Akamai and Fastly in \num{236} \acp{cc}.
There are only \num{11} \acp{cc} that one \ac{as} uniquely covers; in all cases, this \ac{as} is Cloudflare.
\akamaipr covers all \acp{cc} that \akamaiest covers plus \num{212} more.
The analysis of cities covered by subnets (see \Cref{sec:egress-cities}) shows an even distribution across operators (\numrange{800}{1000}) for IPv4.
However, while Fastly does not, Akamai and Cloudflare provide a manifold (\sk{14} and \sk{5} respectively) of cities with IPv6 subnets.
Note that more than \sperc{58} of all subnets cover the US, and the second largest \ac{cc} is DE, with only \sperc{3.6}.
Therefore, \privaterelay has a massive bias towards the US in its current deployment, while a set of \num{123} countries are assigned less than 50 subnets each.

\footnotetext{Data by \copyright OpenStreetMap (http://openstreetmap.org/copyright), under ODbL (http://www.openstreetmap.org/copyright)}
\subsection{Scans Through the Relay}
\label{sec:ingress-egress}

This section looks at the scans through the \privaterelay.
We perform the two requests (using curl and Safari) every five minutes over over a day on multiple days in May 2022.
We look at: \first the chosen egress operator and \second the egress address rotation behavior.

\begin{figure}
    \centering
    \includegraphics{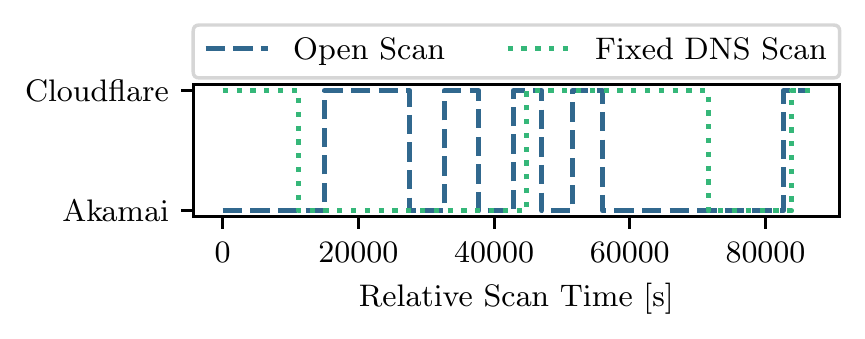}
    \caption{Egress operator changes over the course of a scan day.}
    \label{fig:egress-changes}
\end{figure}

Cloudflare and \akamaipr are the only \acp{as} visible as egress operators.
Fastly's absence is explained by its sparse presence at our measurement location.
\Cref{fig:egress-changes} shows changes between egress operators relative to the scan start.
The open \ac{dns} resolution scan has a handful of seemingly regular operator changes in the middle of its scan, but no pattern over the full scan time is visible.
Similarly, the \emph{fixed} scan does not indicate regular operator changes.

Compared to operators, egress addresses change more regularly, on average, after every second request.
We adapt our scan to reduce the time between each request round to 30 seconds to get a more fine-grained resolution.
During the observation period of 48 hours we find six different egress addresses from four egress subnets.
In more than \sperc{66} of the request attempts, the address changes compared to the previous one.
Due to the relatively low number of six different addresses, it seems plausible that the egress relay selects the address per connection attempt.
We can also support this claim based on the occurrence of different egress addresses for the parallel curl and Safari requests.
To summarize these findings, we can verify the whitepaper's~\cite{privaterelay} claim about address rotation, and it seems suitable to protect the user's IP address from multiple operators in parallel.

Finally, we did not observe egress behavior or address differences when forcing a specific ingress relay address.
\section{Related Work}
\label{sec:related}

\privaterelay has similar goals and architecture as other anonymization tools (\eg \ac{tor}) and it encounters the same problems.
Different research groups \cite{murdoch-tor-traffic, feamster-locating-tor, overlier-locating-servers} showed how \ac{tor} services could be located and passively observed to perform traffic analysis.
Others \cite{edman2009asawareness,johnson2013correlation,nithyanand2016measuring,nasr2018tor} analyzed to which degree \acp{as} can de-anonymize \ac{tor} traffic based on the correlation of traffic entering and leaving the \ac{tor} layers.
Given the small number of involved operators of the relay network and \akamaipr containing ingress and egress relays, similar correlations are drastically more straightforward.
We discuss this in more detail in \Cref{sec:discussion}.

\ac{masque} is used by \privaterelay to proxy the users traffic.
\citeauthor{kuehlewind-masque} \cite{kuehlewind-masque} evaluated the performance metrics of \ac{masque} proxies and found, among other things, an increased RTT when congestion occurs on the target host.

\section{Discussion and Conclusion}
\label{sec:discussion}
Shortly after the announcement of \privaterelay, the system gained significant attention, especially from network operators debating its architecture and potential impact.
This section discusses our results and how research and network operators can use them to prepare for more significant system adoption.
Moreover, we present findings as part of our system analysis, giving a new viewpoint on the promised privacy claims.
In the following, we assume a wide adoption of \privaterelay in the future.

\textbf{Passive Measurements and \privaterelay:}
Network engineers and researchers use passive network measurements to, \eg analyze service usage, traffic categories, and user behavior~\cite{feldmann2021covid,fontugne2017traffic,trevisan2020edge}.
Clients moving seemingly randomly from one egress address to a different one were not yet part of the requirements.
Furthermore, the service potentially multiplexes various traffic and service types in the future.
Especially the egress address rotation and the fact that a client can have multiple parallel connections with differing egress addresses pose a challenge for passive analyses.
These properties differ significantly from the behavior of similar technologies.
Therefore, \privaterelay potentially introduces a new client request pattern that might be classified as anomalous by \acp{ids} (\eg see issue report on Imperva DDoS and \privaterelay~\cite{imperva}).
We suggest consulting the published egress list to identify matching addresses to mitigate the issues.

Research evaluating user behavior using network monitoring in access networks and ISPs needs to tae the client-side properties of the service into account.
The analysis of ISP network data, as done by \citeauthor{trevisan2020edge} \cite{trevisan2020edge} and \citeauthor{feldmann2021covid} \cite{feldmann2021covid}, would not be able to differentiate the service types for relay traffic.
Ingress relays will appear as a highly active destination, but the attribution of traffic to the user's visiting service is impossible.
Moreover, ISPs need to evaluate their paths towards the ingress addresses as an increased load on these should be expected.
These newly appearing traffic patterns can be attributed to our ingress address dataset.

\textbf{Akamai's Presence and its Implications:} \reviewfix{D.4}
Apple claims that tracking users is impossible due to the two layers and separation between operators into distinct entities.
The corresponding stated goal is: ``No one entity can see both who a user is (IP address) and what they are accessing (origin server)''\cite{ietf-slides}.
According to Apple's claims, distributing the service's duties prevents such a correlation.
We use our previous findings to evaluate this claim on a network level.

If an operator can see both the connection by the client to the ingress and the traffic from the egress relay to the target, it can infer the actual communication partners.
The \ac{masque} draft~\cite{masque_draft} explicitly lists traffic analysis as an issue the protocol cannot overcome.
If an observing entity is the egress relay operator, it can use the timing of the requests to extract metadata information and use the relay's provided geohash.
The service derives the geohash from the IP address geolocation, and when the entity observes the client's IP address on the ingress, it can derive its approximate geohashes.

\akamaipr locates the largest number of ingress as well as egress relays.
Therefore, it was no surprise that we find occurrences of the \ac{as} in the ingress and the egress in our relay scan (see \Cref{sec:ingress-egress}).
We validated these findings through traceroute measurements and found the same last hop address for ingress and egress addresses.
In contrast to the \ac{masque} draft, \privaterelay does not only define the protocol but has also designed its architecture and deployment strategy.
Apple could ensure that ingress and egress addresses are not part of the same \ac{as} and entity to prevent such an issue in the future.
Nevertheless, currently, the \ac{as} contributing the largest share of ingress and egress relays is also the source of this traffic analysis issue.

\akamaipr as the culprit of this problem has only one publicly visible peering link to \akamaiest{}, providing egress relay nodes.
In total, the \num{478} IPv4 and \num{1335} IPv6 \ac{bgp} prefixes are visible.
We find at least one ingress relay in 201 and one egress relay in \num{1472} prefixes (IPv4+IPv6).
Given that ingress and egress relays at least do not share the same BGP prefix, \sperc{92.2} (\num{1673}) of announced prefixes are used for \privaterelay.

We examined the BGP visibility of the AS monthly from 2016 to 2022.
Its first occurrence was detected in June 2021, coinciding with the launch of \privaterelay.
This information strongly hints that \akamaipr is specifically used for \privaterelay.
For this purpose, we have sent an inquiry to Akamai, which could not be answered due to proprietary information.
Nevertheless, it seems to be an odd decision to design the system with the claimed privacy goals, but these mentioned issues as Apple is directly involved in writing the \ac{masque} draft.

\textbf{Conclusion:}
In this work, we provide an overview of \privaterelay and highlight its effect on future research and network operators.
We collect an ingress address data set that can be used to identify client connections to the \privaterelay.
Our scans uncover the Akamai private relay \ac{as} that locates ingress and egress relays and potentially allows traffic correlation.

During this study, we found some research topics we consider important in the future:
\first Where and how is traffic routed to and from the relay nodes? Does the system have bottlenecks that can lead to congestion for its users?
\second How does the system evolve, and where is it available? Some countries disallow its usage due to the censorship evasion possibility.
\third How does the service impact the user's QoE? Apple claims the impact is low, and caching would also lead to faster page load times.

\section{Ethics}
\label{sec:ethics}

Before conducting active measurements we follow an internal multiparty approval process, which incorporates proposals by \citeauthor{menloreport} and \citeauthor{partridge2016ethical}.
We asses if our measurements induce harm on scanned infrastructure or other actors and minimize the risk of doing so.
In this paper we performed active Internet scans towards a limited target set.
We apply a strict query rate limit for all of our scans.
Our scanning IP range has an according WHOIS entry, reverse \ac{dns} entries and the scanning hosts operate an informational webpage.
We comply with opt-out requests and operate a blocklist for all our scans.
All our active scans target either services or our own infrastructure.
This works measurements did not trigger any abuse notification.

We apply additional measures to reduce the number of \ac{ecs} queries.
To lower the load on the authoritative name servers, we sparsely scan the address space not seen as routable by our local BGP router.
This address space is not valuable as it does not contain possible users for the service.
We also respect the \ac{ecs} information sent by the name server, \ie the validity of the answer for a prefix of any size.
If the prefix is large than /24, we do not perform any other \ac{ecs} query within that prefix.
These measures help to reduce the number of redundant queries needed and therefore are beneficial for the scanned infrastructure.

\label{body}

\begin{acks}
	The authors would like to thank the anonymous reviewers and our
	shepherd Alan Mislove for their valuable feedback. This work
	was partially funded by the German Federal Ministry of Education
	and Research under the project PRIMEnet, grant 16KIS1370.
\end{acks}

\balance

\bibliographystyle{ACM-Reference-Format}
\bibliography{lit}


\begin{thebibliography}{39}


\ifx \showCODEN    \undefined \def \showCODEN     #1{\unskip}     \fi
\ifx \showDOI      \undefined \def \showDOI       #1{#1}\fi
\ifx \showISBNx    \undefined \def \showISBNx     #1{\unskip}     \fi
\ifx \showISBNxiii \undefined \def \showISBNxiii  #1{\unskip}     \fi
\ifx \showISSN     \undefined \def \showISSN      #1{\unskip}     \fi
\ifx \showLCCN     \undefined \def \showLCCN      #1{\unskip}     \fi
\ifx \shownote     \undefined \def \shownote      #1{#1}          \fi
\ifx \showarticletitle \undefined \def \showarticletitle #1{#1}   \fi
\ifx \showURL      \undefined \def \showURL       {\relax}        \fi
\providecommand\bibfield[2]{#2}
\providecommand\bibinfo[2]{#2}
\providecommand\natexlab[1]{#1}
\providecommand\showeprint[2][]{arXiv:#2}

\bibitem[\protect\citeauthoryear{??}{mas}{2020}]%
        {masque_wg}
 \bibinfo{year}{2020}\natexlab{}.
\newblock \bibinfo{booktitle}{\emph{{Multiplexed Application Substrate over
  QUIC Encryption}}}.
\newblock
\urldef\tempurl%
\url{https://datatracker.ietf.org/doc/charter-ietf-masque/}
\showURL{%
Retrieved May 11, 2022 from \tempurl}


\bibitem[\protect\citeauthoryear{??}{qua}{2022}]%
        {quad9}
 \bibinfo{year}{2022}\natexlab{}.
\newblock \bibinfo{booktitle}{\emph{{Quad9}}}.
\newblock
\urldef\tempurl%
\url{https://www.quad9.net/}
\showURL{%
Retrieved 2022-05-14 from \tempurl}


\bibitem[\protect\citeauthoryear{{APNIC}}{{APNIC}}{2022}]%
        {aspop}
\bibfield{author}{\bibinfo{person}{{APNIC}}.} \bibinfo{year}{2022}\natexlab{}.
\newblock \bibinfo{booktitle}{\emph{{Visible ASNs: Customer Populations}}}.
\newblock
\urldef\tempurl%
\url{https://stats.labs.apnic.net/aspop}
\showURL{%
Retrieved 2022-05-14 from \tempurl}


\bibitem[\protect\citeauthoryear{Bailey, Dittrich, Kenneally, and
  Maughan}{Bailey et~al\mbox{.}}{2012}]%
        {menloreport}
\bibfield{author}{\bibinfo{person}{Michael Bailey}, \bibinfo{person}{David
  Dittrich}, \bibinfo{person}{Erin Kenneally}, {and} \bibinfo{person}{Doug
  Maughan}.} \bibinfo{year}{2012}\natexlab{}.
\newblock \showarticletitle{The Menlo Report}.
\newblock \bibinfo{journal}{\emph{IEEE Security Privacy}}
  (\bibinfo{year}{2012}).
\newblock


\bibitem[\protect\citeauthoryear{Bajpai, Eravuchira, and
  Sch\"{o}nw\"{a}lder}{Bajpai et~al\mbox{.}}{2015}]%
        {bajpai2015atlas}
\bibfield{author}{\bibinfo{person}{Vaibhav Bajpai},
  \bibinfo{person}{Steffie~Jacob Eravuchira}, {and} \bibinfo{person}{J\"{u}rgen
  Sch\"{o}nw\"{a}lder}.} \bibinfo{year}{2015}\natexlab{}.
\newblock \showarticletitle{Lessons Learned From Using the RIPE Atlas Platform
  for Measurement Research}.
\newblock \bibinfo{journal}{\emph{SIGCOMM Comput. Commun. Rev.}}
  (\bibinfo{date}{jul} \bibinfo{year}{2015}).
\newblock


\bibitem[\protect\citeauthoryear{Calder, Fan, Hu, Katz-Bassett, Heidemann, and
  Govindan}{Calder et~al\mbox{.}}{2013}]%
        {calder2013ecs}
\bibfield{author}{\bibinfo{person}{Matt Calder}, \bibinfo{person}{Xun Fan},
  \bibinfo{person}{Zi Hu}, \bibinfo{person}{Ethan Katz-Bassett},
  \bibinfo{person}{John Heidemann}, {and} \bibinfo{person}{Ramesh Govindan}.}
  \bibinfo{year}{2013}\natexlab{}.
\newblock \showarticletitle{{Mapping the Expansion of Google's Serving
  Infrastructure}}. In \bibinfo{booktitle}{\emph{Proceedings of the 2013
  Conference on Internet Measurement Conference}} (Barcelona, Spain)
  \emph{(\bibinfo{series}{IMC '13})}. \bibinfo{publisher}{Association for
  Computing Machinery}, \bibinfo{address}{New York, NY, USA}.
\newblock


\bibitem[\protect\citeauthoryear{{Cisco}}{{Cisco}}{2022}]%
        {opendns}
\bibfield{author}{\bibinfo{person}{{Cisco}}.} \bibinfo{year}{2022}\natexlab{}.
\newblock \bibinfo{booktitle}{\emph{{OpenDNS}}}.
\newblock
\urldef\tempurl%
\url{https://www.opendns.com/}
\showURL{%
Retrieved 2022-05-14 from \tempurl}


\bibitem[\protect\citeauthoryear{{Cloudflare}}{{Cloudflare}}{2022}]%
        {cloudflare-dns}
\bibfield{author}{\bibinfo{person}{{Cloudflare}}.}
  \bibinfo{year}{2022}\natexlab{}.
\newblock \bibinfo{booktitle}{\emph{{1.1.1.1}}}.
\newblock
\urldef\tempurl%
\url{https://developers.cloudflare.com/1.1.1.1/}
\showURL{%
Retrieved 2022-05-14 from \tempurl}


\bibitem[\protect\citeauthoryear{Contavalli, van~der Gaast, Lawrence, and
  Kumari}{Contavalli et~al\mbox{.}}{2016}]%
        {rfc7871}
\bibfield{author}{\bibinfo{person}{Carlo Contavalli}, \bibinfo{person}{Wilmer
  van~der Gaast}, \bibinfo{person}{David~C Lawrence}, {and}
  \bibinfo{person}{Warren~"Ace" Kumari}.} \bibinfo{year}{2016}\natexlab{}.
\newblock \bibinfo{title}{{Client Subnet in DNS Queries}}.
\newblock \bibinfo{howpublished}{RFC 7871}.
\newblock
\urldef\tempurl%
\url{https://www.rfc-editor.org/info/rfc7871}
\showURL{%
\tempurl}


\bibitem[\protect\citeauthoryear{Edman and Syverson}{Edman and
  Syverson}{2009}]%
        {edman2009asawareness}
\bibfield{author}{\bibinfo{person}{Matthew Edman} {and} \bibinfo{person}{Paul
  Syverson}.} \bibinfo{year}{2009}\natexlab{}.
\newblock \showarticletitle{{As-Awareness in Tor Path Selection}}. In
  \bibinfo{booktitle}{\emph{Proc. ACM SIGSAC Conference on Computer and
  Communications Security (CCS)}} (Chicago, Illinois, USA).
\newblock


\bibitem[\protect\citeauthoryear{Feamster and Dingledine}{Feamster and
  Dingledine}{2004}]%
        {feamster-locating-tor}
\bibfield{author}{\bibinfo{person}{Nick Feamster} {and} \bibinfo{person}{Roger
  Dingledine}.} \bibinfo{year}{2004}\natexlab{}.
\newblock \showarticletitle{{Location Diversity in Anonymity Networks}}. In
  \bibinfo{booktitle}{\emph{Proceedings of the 2004 ACM Workshop on Privacy in
  the Electronic Society}} (Washington DC, USA).
\newblock


\bibitem[\protect\citeauthoryear{Feldmann, Gasser, Lichtblau, Pujol, Poese,
  Dietzel, Wagner, Wichtlhuber, Tapiador, Vallina-Rodriguez, Hohlfeld, and
  Smaragdakis}{Feldmann et~al\mbox{.}}{2020}]%
        {feldmann2021covid}
\bibfield{author}{\bibinfo{person}{Anja Feldmann}, \bibinfo{person}{Oliver
  Gasser}, \bibinfo{person}{Franziska Lichtblau}, \bibinfo{person}{Enric
  Pujol}, \bibinfo{person}{Ingmar Poese}, \bibinfo{person}{Christoph Dietzel},
  \bibinfo{person}{Daniel Wagner}, \bibinfo{person}{Matthias Wichtlhuber},
  \bibinfo{person}{Juan Tapiador}, \bibinfo{person}{Narseo Vallina-Rodriguez},
  \bibinfo{person}{Oliver Hohlfeld}, {and} \bibinfo{person}{Georgios
  Smaragdakis}.} \bibinfo{year}{2020}\natexlab{}.
\newblock \showarticletitle{{The Lockdown Effect: Implications of the COVID-19
  Pandemic on Internet Traffic}}. In \bibinfo{booktitle}{\emph{Proc. ACM Int.
  Measurement Conference (IMC)}} (Virtual Event, USA).
\newblock


\bibitem[\protect\citeauthoryear{Fontugne, Abry, Fukuda, Veitch, Cho, Borgnat,
  and Wendt}{Fontugne et~al\mbox{.}}{2017}]%
        {fontugne2017traffic}
\bibfield{author}{\bibinfo{person}{Romain Fontugne}, \bibinfo{person}{Patrice
  Abry}, \bibinfo{person}{Kensuke Fukuda}, \bibinfo{person}{Darryl Veitch},
  \bibinfo{person}{Kenjiro Cho}, \bibinfo{person}{Pierre Borgnat}, {and}
  \bibinfo{person}{Herwig Wendt}.} \bibinfo{year}{2017}\natexlab{}.
\newblock \showarticletitle{{Scaling in Internet Traffic: A 14 Year and 3 Day
  Longitudinal Study, With Multiscale Analyses and Random Projections}}.
\newblock \bibinfo{journal}{\emph{IEEE/ACM Transactions on Networking}}
  (\bibinfo{year}{2017}).
\newblock


\bibitem[\protect\citeauthoryear{{Google}}{{Google}}{2022}]%
        {google-pub-dns}
\bibfield{author}{\bibinfo{person}{{Google}}.} \bibinfo{year}{2022}\natexlab{}.
\newblock \bibinfo{booktitle}{\emph{{Public DNS}}}.
\newblock
\urldef\tempurl%
\url{https://developers.google.com/speed/public-dns/}
\showURL{%
Retrieved 2022-05-14 from \tempurl}


\bibitem[\protect\citeauthoryear{Inc}{Inc}{2021a}]%
        {egress-nodes}
\bibfield{author}{\bibinfo{person}{Apple Inc}.}
  \bibinfo{year}{2021}\natexlab{a}.
\newblock \bibinfo{booktitle}{\emph{{Access IP geolocation feeds}}}.
\newblock
\urldef\tempurl%
\url{https://mask-api.icloud.com/egress-ip-ranges.csv}
\showURL{%
Retrieved 2022-05-16 from \tempurl}


\bibitem[\protect\citeauthoryear{Inc}{Inc}{2021b}]%
        {privaterelay}
\bibfield{author}{\bibinfo{person}{Apple Inc}.}
  \bibinfo{year}{2021}\natexlab{b}.
\newblock \showarticletitle{{iCloud Private Relay Overview}}.
\newblock  (\bibinfo{year}{2021}).
\newblock
\urldef\tempurl%
\url{https://www.apple.com/privacy/docs/iCloud_Private_Relay_Overview_Dec2021.PDF}
\showURL{%
\tempurl}


\bibitem[\protect\citeauthoryear{Johnson, Wacek, Jansen, Sherr, and
  Syverson}{Johnson et~al\mbox{.}}{2013a}]%
        {johnson-tor-routing}
\bibfield{author}{\bibinfo{person}{Aaron Johnson}, \bibinfo{person}{Chris
  Wacek}, \bibinfo{person}{Rob Jansen}, \bibinfo{person}{Micah Sherr}, {and}
  \bibinfo{person}{Paul Syverson}.} \bibinfo{year}{2013}\natexlab{a}.
\newblock \showarticletitle{{Users Get Routed: Traffic Correlation on Tor by
  Realistic Adversaries}}. In \bibinfo{booktitle}{\emph{Proc. ACM SIGSAC
  Conference on Computer and Communications Security (CCS)}} (Berlin, Germany).
\newblock


\bibitem[\protect\citeauthoryear{Johnson, Wacek, Jansen, Sherr, and
  Syverson}{Johnson et~al\mbox{.}}{2013b}]%
        {johnson2013correlation}
\bibfield{author}{\bibinfo{person}{Aaron Johnson}, \bibinfo{person}{Chris
  Wacek}, \bibinfo{person}{Rob Jansen}, \bibinfo{person}{Micah Sherr}, {and}
  \bibinfo{person}{Paul Syverson}.} \bibinfo{year}{2013}\natexlab{b}.
\newblock \showarticletitle{{Users Get Routed: Traffic Correlation on Tor by
  Realistic Adversaries}}. In \bibinfo{booktitle}{\emph{Proc. ACM SIGSAC
  Conference on Computer and Communications Security (CCS)}} (Berlin, Germany).
\newblock


\bibitem[\protect\citeauthoryear{K\"{u}hlewind, Carlander-Reuterfelt, Ihlar,
  and Westerlund}{K\"{u}hlewind et~al\mbox{.}}{2021}]%
        {kuehlewind-masque}
\bibfield{author}{\bibinfo{person}{Mirja K\"{u}hlewind},
  \bibinfo{person}{Matias Carlander-Reuterfelt}, \bibinfo{person}{Marcus
  Ihlar}, {and} \bibinfo{person}{Magnus Westerlund}.}
  \bibinfo{year}{2021}\natexlab{}.
\newblock \showarticletitle{Evaluation of QUIC-Based MASQUE Proxying}. In
  \bibinfo{booktitle}{\emph{Proceedings of the 2021 Workshop on Evolution,
  Performance and Interoperability of QUIC}} (Virtual Event, Germany).
\newblock


\bibitem[\protect\citeauthoryear{Lalkaka}{Lalkaka}{2017}]%
        {argo}
\bibfield{author}{\bibinfo{person}{Rustam Lalkaka}.}
  \bibinfo{year}{2017}\natexlab{}.
\newblock \bibinfo{booktitle}{\emph{{Introducing Argo — A faster, more
  reliable, more secure Internet for everyone}}}.
\newblock
\urldef\tempurl%
\url{https://blog.cloudflare.com/argo/}
\showURL{%
Retrieved 2022-09-13 from \tempurl}


\bibitem[\protect\citeauthoryear{Lalkaka}{Lalkaka}{2022}]%
        {cf-icpr}
\bibfield{author}{\bibinfo{person}{Rustam Lalkaka}.}
  \bibinfo{year}{2022}\natexlab{}.
\newblock \bibinfo{booktitle}{\emph{{iCloud Private Relay: information for
  Cloudflare customers}}}.
\newblock
\urldef\tempurl%
\url{https://blog.cloudflare.com/icloud-private-relay/}
\showURL{%
Retrieved 2022-09-13 from \tempurl}


\bibitem[\protect\citeauthoryear{Murdoch and Danezis}{Murdoch and
  Danezis}{2005}]%
        {murdoch-tor-traffic}
\bibfield{author}{\bibinfo{person}{Steven~J. Murdoch} {and}
  \bibinfo{person}{George Danezis}.} \bibinfo{year}{2005}\natexlab{}.
\newblock \showarticletitle{{Low-Cost Traffic Analysis of Tor}}. In
  \bibinfo{booktitle}{\emph{Proc. IEEE Symposium on Security and Privacy
  (S\&P)}}.
\newblock


\bibitem[\protect\citeauthoryear{Nasr, Bahramali, and Houmansadr}{Nasr
  et~al\mbox{.}}{2018}]%
        {nasr2018tor}
\bibfield{author}{\bibinfo{person}{Milad Nasr}, \bibinfo{person}{Alireza
  Bahramali}, {and} \bibinfo{person}{Amir Houmansadr}.}
  \bibinfo{year}{2018}\natexlab{}.
\newblock \showarticletitle{{DeepCorr: Strong Flow Correlation Attacks on Tor
  Using Deep Learning}}. In \bibinfo{booktitle}{\emph{Proc. ACM SIGSAC
  Conference on Computer and Communications Security (CCS)}} (Toronto, Canada).
\newblock


\bibitem[\protect\citeauthoryear{Nellis and Dave}{Nellis and Dave}{2022}]%
        {prnotavailablechina}
\bibfield{author}{\bibinfo{person}{Stephen Nellis} {and}
  \bibinfo{person}{Paresh Dave}.} \bibinfo{year}{2022}\natexlab{}.
\newblock \bibinfo{booktitle}{\emph{{Apple's new 'private relay' feature will
  not be available in China}}}.
\newblock
\urldef\tempurl%
\url{https://www.reuters.com/world/china/apples-new-private-relay-feature-will-not-be-available-china-2021-06-07/}
\showURL{%
Retrieved 2022-09-03 from \tempurl}


\bibitem[\protect\citeauthoryear{Nerenberg}{Nerenberg}{2022}]%
        {imperva}
\bibfield{author}{\bibinfo{person}{Lyndon Nerenberg}.}
  \bibinfo{year}{2022}\natexlab{}.
\newblock \bibinfo{booktitle}{\emph{{Imperva / Apple Private Relay issues}}}.
\newblock
\urldef\tempurl%
\url{https://mailman.nanog.org/pipermail/nanog/2022-September/220491.html}
\showURL{%
Retrieved 2022-09-15 from \tempurl}


\bibitem[\protect\citeauthoryear{Nithyanand, Starov, Zair, Gill, and
  Schapira}{Nithyanand et~al\mbox{.}}{2016}]%
        {nithyanand2016measuring}
\bibfield{author}{\bibinfo{person}{Rishab Nithyanand}, \bibinfo{person}{Oleksii
  Starov}, \bibinfo{person}{Adva Zair}, \bibinfo{person}{Phillipa Gill}, {and}
  \bibinfo{person}{Michael Schapira}.} \bibinfo{year}{2016}\natexlab{}.
\newblock \showarticletitle{{Measuring and Mitigating AS-level Adversaries
  Against Tor}}. In \bibinfo{booktitle}{\emph{Proc. Network and Distributed
  System Security Symposium (NDSS)}}. \bibinfo{address}{San Diego, CA}.
\newblock


\bibitem[\protect\citeauthoryear{Overlier and Syverson}{Overlier and
  Syverson}{2006}]%
        {overlier-locating-servers}
\bibfield{author}{\bibinfo{person}{Lasse Overlier} {and} \bibinfo{person}{Paul
  Syverson}.} \bibinfo{year}{2006}\natexlab{}.
\newblock \showarticletitle{{Locating Hidden Servers}}. In
  \bibinfo{booktitle}{\emph{Proc. IEEE Symposium on Security and Privacy
  (S\&P)}}.
\newblock


\bibitem[\protect\citeauthoryear{Partridge and Allman}{Partridge and
  Allman}{2016}]%
        {partridge2016ethical}
\bibfield{author}{\bibinfo{person}{Craig Partridge} {and} \bibinfo{person}{Mark
  Allman}.} \bibinfo{year}{2016}\natexlab{}.
\newblock \showarticletitle{Ethical Considerations in Network Measurement
  Papers}.
\newblock \bibinfo{journal}{\emph{Commun. ACM}} (\bibinfo{year}{2016}).
\newblock


\bibitem[\protect\citeauthoryear{Pauly and Schinazi}{Pauly and
  Schinazi}{2022}]%
        {masque_draft}
\bibfield{author}{\bibinfo{person}{Tommy Pauly} {and} \bibinfo{person}{David
  Schinazi}.} \bibinfo{year}{2022}\natexlab{}.
\newblock \bibinfo{booktitle}{\emph{{QUIC-Aware Proxying Using HTTP}}}.
\newblock \bibinfo{type}{Internet-Draft} draft-pauly-masque-quic-proxy-03.
  \bibinfo{institution}{Internet Engineering Task Force}.
\newblock
\urldef\tempurl%
\url{https://datatracker.ietf.org/doc/html/draft-pauly-masque-quic-proxy-03}
\showURL{%
\tempurl}
\newblock
\shownote{Work in Progress}.


\bibitem[\protect\citeauthoryear{{Pauly, Tommy}}{{Pauly, Tommy}}{2008}]%
        {ietf-slides}
\bibfield{author}{\bibinfo{person}{{Pauly, Tommy}}.}
  \bibinfo{year}{2008}\natexlab{}.
\newblock \bibinfo{booktitle}{\emph{{iCloud Private Relay}}}.
\newblock
\urldef\tempurl%
\url{https://datatracker.ietf.org/meeting/111/materials/slides-111-pearg-private-relay-00}
\showURL{%
Retrieved May 11, 2022 from \tempurl}


\bibitem[\protect\citeauthoryear{Sattler, Aulbach, Zirngibl, and Carle}{Sattler
  et~al\mbox{.}}{2022}]%
        {datatum}
\bibfield{author}{\bibinfo{person}{Patrick Sattler}, \bibinfo{person}{Juliane
  Aulbach}, \bibinfo{person}{Johannes Zirngibl}, {and} \bibinfo{person}{Georg
  Carle}.} \bibinfo{year}{2022}\natexlab{}.
\newblock \bibinfo{booktitle}{\emph{{Data and Analysis at TUM University
  Library}}}.
\newblock
\urldef\tempurl%
\url{https://mediatum.ub.tum.de/1687050}
\showURL{%
\tempurl}
\newblock
\shownote{doi:10.14459/2022mp1687050}.


\bibitem[\protect\citeauthoryear{Schinazi}{Schinazi}{2022}]%
        {draft-ietf-masque-connect-udp}
\bibfield{author}{\bibinfo{person}{David Schinazi}.}
  \bibinfo{year}{2022}\natexlab{}.
\newblock \bibinfo{title}{{Proxying UDP in HTTP}}.
\newblock \bibinfo{howpublished}{RFC 9298}.
\newblock
\urldef\tempurl%
\url{https://www.rfc-editor.org/info/rfc9298}
\showURL{%
\tempurl}


\bibitem[\protect\citeauthoryear{Statista}{Statista}{2022}]%
        {phonemarketshare}
\bibfield{author}{\bibinfo{person}{Statista}.} \bibinfo{year}{2022}\natexlab{}.
\newblock \bibinfo{booktitle}{\emph{{Mobile operating systems' market share
  worldwide from January 2012 to January 2022}}}.
\newblock
\urldef\tempurl%
\url{https://www.statista.com/statistics/272698/global-market-share-held-by-mobile-operating-systems-since-2009/}
\showURL{%
Retrieved 2022-05-14 from \tempurl}


\bibitem[\protect\citeauthoryear{Streibelt, B\"{o}ttger, Chatzis, Smaragdakis,
  and Feldmann}{Streibelt et~al\mbox{.}}{2013}]%
        {streibelt-ecs}
\bibfield{author}{\bibinfo{person}{Florian Streibelt}, \bibinfo{person}{Jan
  B\"{o}ttger}, \bibinfo{person}{Nikolaos Chatzis}, \bibinfo{person}{Georgios
  Smaragdakis}, {and} \bibinfo{person}{Anja Feldmann}.}
  \bibinfo{year}{2013}\natexlab{}.
\newblock \showarticletitle{{Exploring EDNS-Client-Subnet Adopters in Your Free
  Time}}. In \bibinfo{booktitle}{\emph{Proc. ACM Int. Measurement Conference
  (IMC)}} (Barcelona, Spain).
\newblock


\bibitem[\protect\citeauthoryear{Sun, Edmundson, Vanbever, Li, Rexford, Chiang,
  and Mittal}{Sun et~al\mbox{.}}{2015}]%
        {sun-raptor}
\bibfield{author}{\bibinfo{person}{Yixin Sun}, \bibinfo{person}{Anne
  Edmundson}, \bibinfo{person}{Laurent Vanbever}, \bibinfo{person}{Oscar Li},
  \bibinfo{person}{Jennifer Rexford}, \bibinfo{person}{Mung Chiang}, {and}
  \bibinfo{person}{Prateek Mittal}.} \bibinfo{year}{2015}\natexlab{}.
\newblock \showarticletitle{{RAPTOR: Routing Attacks on Privacy in Tor}}. In
  \bibinfo{booktitle}{\emph{24th USENIX Security Symposium (USENIX Security
  15)}}. \bibinfo{publisher}{USENIX Association}, \bibinfo{address}{Washington,
  D.C.}
\newblock


\bibitem[\protect\citeauthoryear{Trevisan, Giordano, Drago, Munafò, and
  Mellia}{Trevisan et~al\mbox{.}}{2020}]%
        {trevisan2020edge}
\bibfield{author}{\bibinfo{person}{Martino Trevisan}, \bibinfo{person}{Danilo
  Giordano}, \bibinfo{person}{Idilio Drago}, \bibinfo{person}{Maurizio~Matteo
  Munafò}, {and} \bibinfo{person}{Marco Mellia}.}
  \bibinfo{year}{2020}\natexlab{}.
\newblock \showarticletitle{{Five Years at the Edge: Watching Internet From the
  ISP Network}}.
\newblock \bibinfo{journal}{\emph{IEEE/ACM Transactions on Networking}}
  (\bibinfo{year}{2020}).
\newblock


\bibitem[\protect\citeauthoryear{{Woods, Ben and Titcomb, James}}{{Woods, Ben
  and Titcomb, James}}{2022}]%
        {telegraph-isps-whining}
\bibfield{author}{\bibinfo{person}{{Woods, Ben and Titcomb, James}}.}
  \bibinfo{year}{2022}\natexlab{}.
\newblock \bibinfo{booktitle}{\emph{{Apple under fire over iPhone encryption
  tech}}}.
\newblock
\urldef\tempurl%
\url{https://www.telegraph.co.uk/business/2022/01/09/apple-fire-iphone-encryption-tech/}
\showURL{%
Retrieved May 11, 2022 from \tempurl}


\bibitem[\protect\citeauthoryear{Wouters, Tschofenig, Gilmore, Weiler, and
  Kivinen}{Wouters et~al\mbox{.}}{2014}]%
        {rfc7250}
\bibfield{author}{\bibinfo{person}{Paul Wouters}, \bibinfo{person}{Hannes
  Tschofenig}, \bibinfo{person}{John~IETF Gilmore}, \bibinfo{person}{Samuel
  Weiler}, {and} \bibinfo{person}{Tero Kivinen}.}
  \bibinfo{year}{2014}\natexlab{}.
\newblock \bibinfo{title}{{Using Raw Public Keys in Transport Layer Security
  (TLS) and Datagram Transport Layer Security (DTLS)}}.
\newblock \bibinfo{howpublished}{RFC 7250}.
\newblock
\urldef\tempurl%
\url{https://doi.org/10.17487/RFC7250}
\showDOI{\tempurl}


\bibitem[\protect\citeauthoryear{Zirngibl, Buschmann, Sattler, Jaeger, Aulbach,
  and Carle}{Zirngibl et~al\mbox{.}}{2021}]%
        {zirngibl2021over9000}
\bibfield{author}{\bibinfo{person}{Johannes Zirngibl},
  \bibinfo{person}{Philippe Buschmann}, \bibinfo{person}{Patrick Sattler},
  \bibinfo{person}{Benedikt Jaeger}, \bibinfo{person}{Juliane Aulbach}, {and}
  \bibinfo{person}{Georg Carle}.} \bibinfo{year}{2021}\natexlab{}.
\newblock \showarticletitle{{It's over 9000: Analyzing early QUIC Deployments
  with the Standardization on the Horizon}}. In \bibinfo{booktitle}{\emph{Proc.
  ACM Int. Measurement Conference (IMC)}} (Virtual Event, USA).
\newblock


\end{thebibliography}
\appendix
\section{Locations covered by Egress Operators}
\label{sec:egress-cities}

\begin{figure*}[h]
	\begin{subfigure}{.5\textwidth}
		\centering
		\includegraphics[width=.9\linewidth]{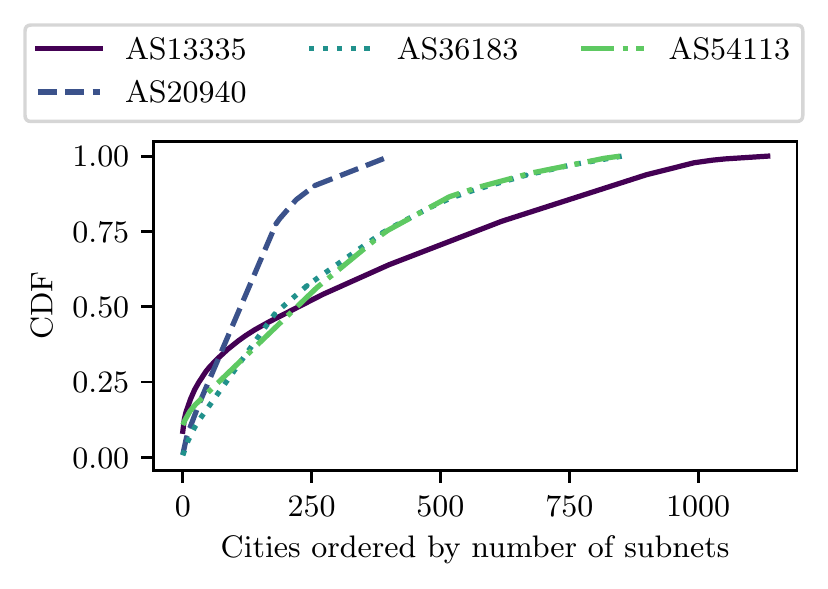}
		\caption{IPv4 - Cities}
		\label{fig:citiesv4cdf}
	\end{subfigure}%
	\begin{subfigure}{.5\textwidth}
		\centering
		\includegraphics[width=.9\linewidth]{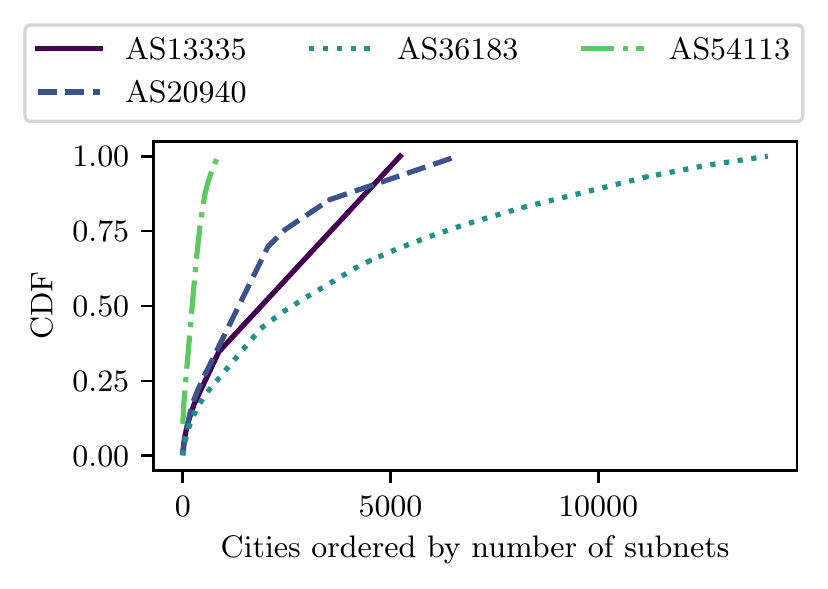}
		\caption{IPv6 - Cities}
		\label{fig:citiesv6cdf}
	\end{subfigure}
	\begin{subfigure}{.5\textwidth}
		\centering
		\includegraphics[width=.9\linewidth]{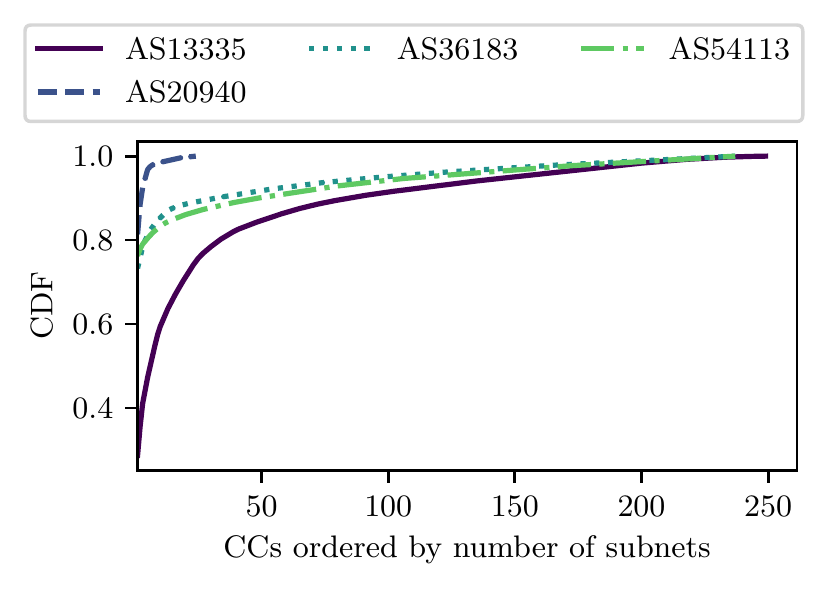}
		\caption{IPv4 - Countries}
		\label{fig:ccv4cdf}
	\end{subfigure}%
	\begin{subfigure}{.5\textwidth}
		\centering
		\includegraphics[width=.9\linewidth]{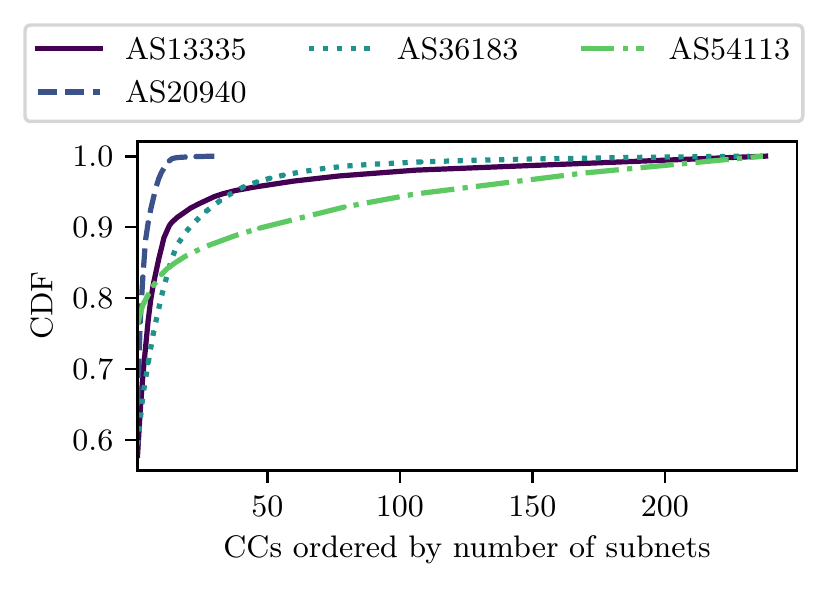}
		\caption{IPv6 - Countries}
		\label{fig:ccv6cdf}
	\end{subfigure}
	\caption{Distribution of subnet locations across countries and cities per egress operator \ac{as}.}
	\label{fig:cccdf}
\end{figure*}

\begin{figure*}
    \begin{subfigure}{.33\textwidth}
        \centering
        \includegraphics[width=.9\linewidth]{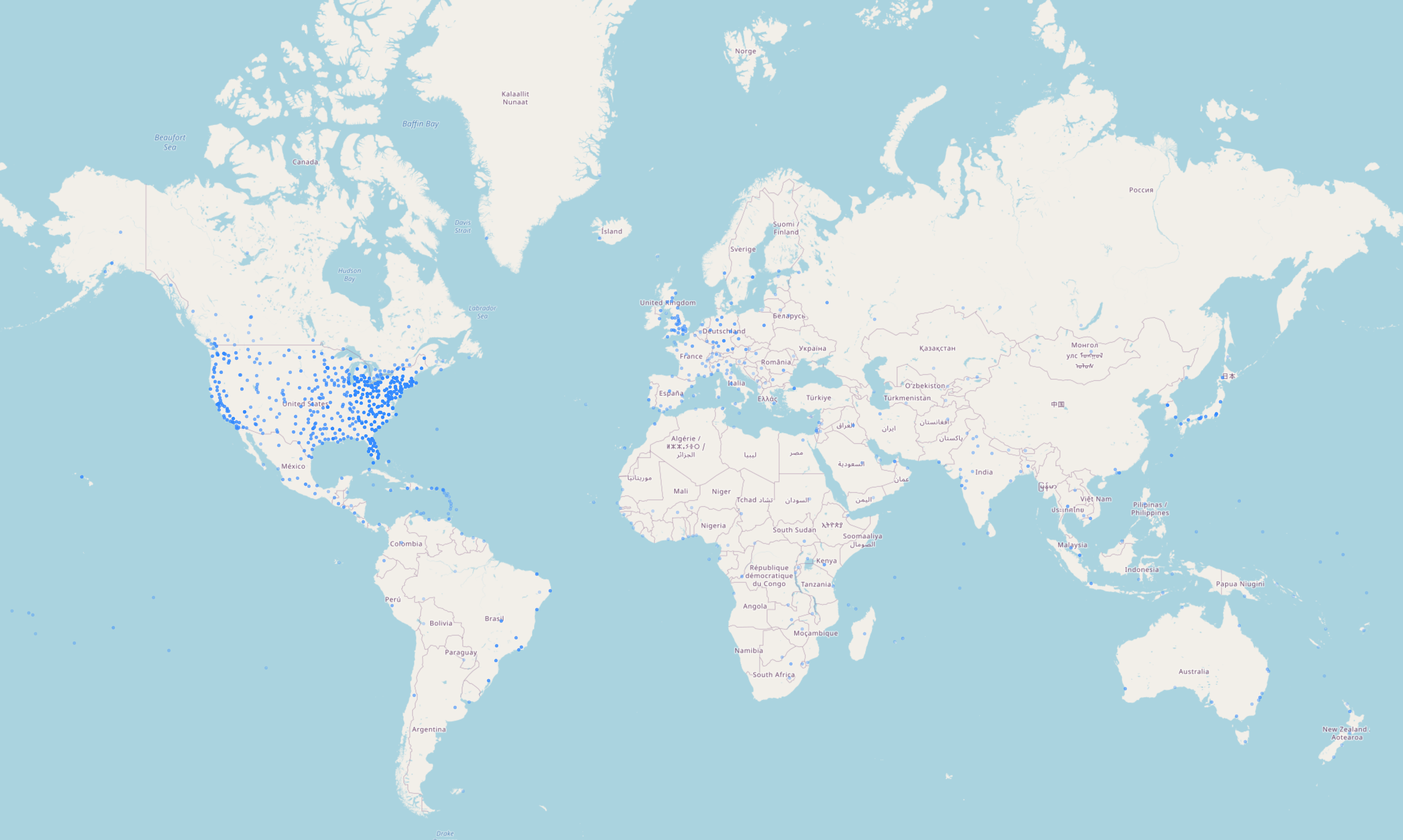}
        \caption{\akamaipr and \akamaiest (IPv4)}
        \label{fig:sub1_v4}
    \end{subfigure}
    \begin{subfigure}{.33\textwidth}
        \centering
        \includegraphics[width=.9\linewidth]{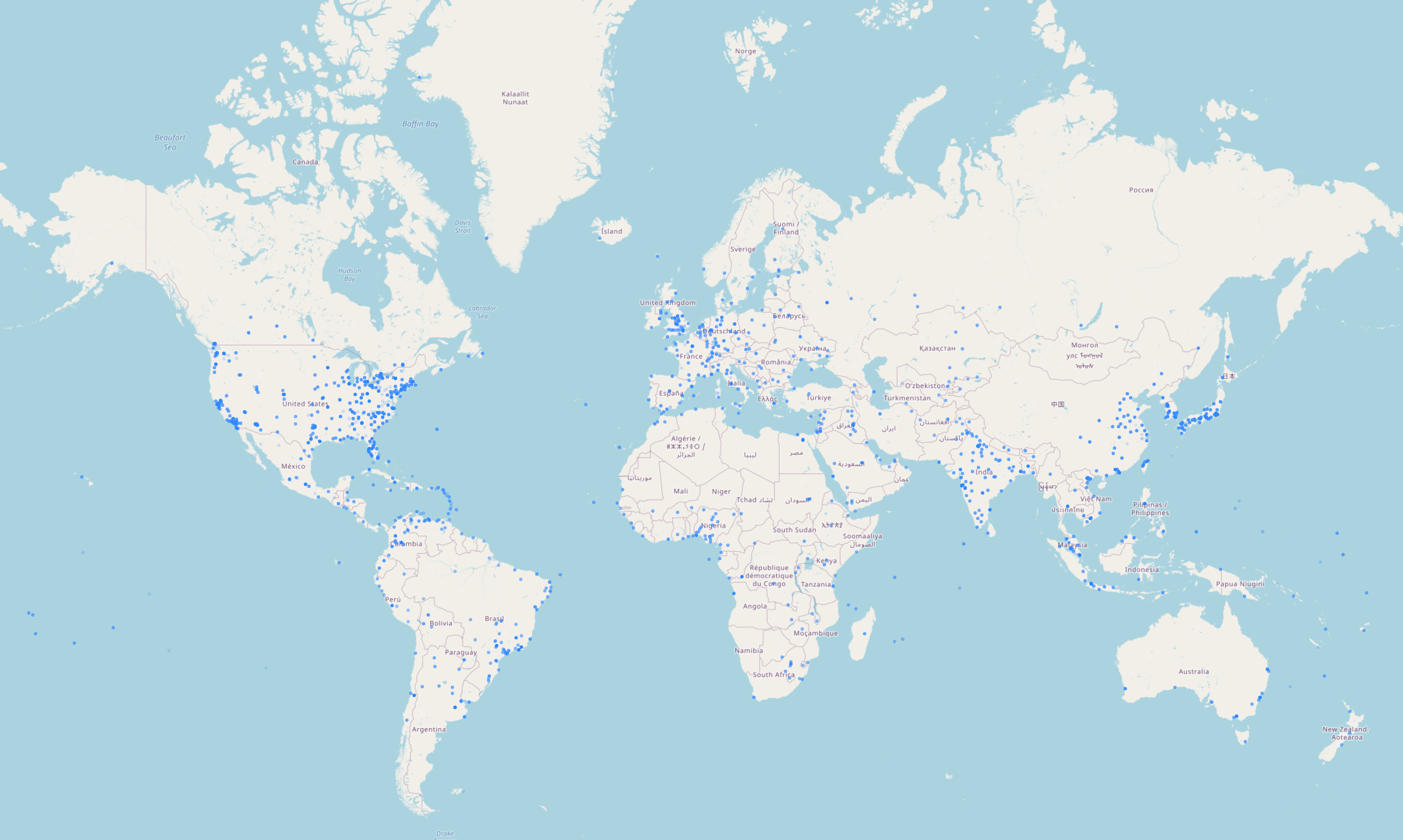}
        \caption{Cloudflare (IPv4)}
        \label{fig:sub2_v4}
    \end{subfigure}
    \begin{subfigure}{.33\textwidth}
        \centering
        \includegraphics[width=.9\linewidth]{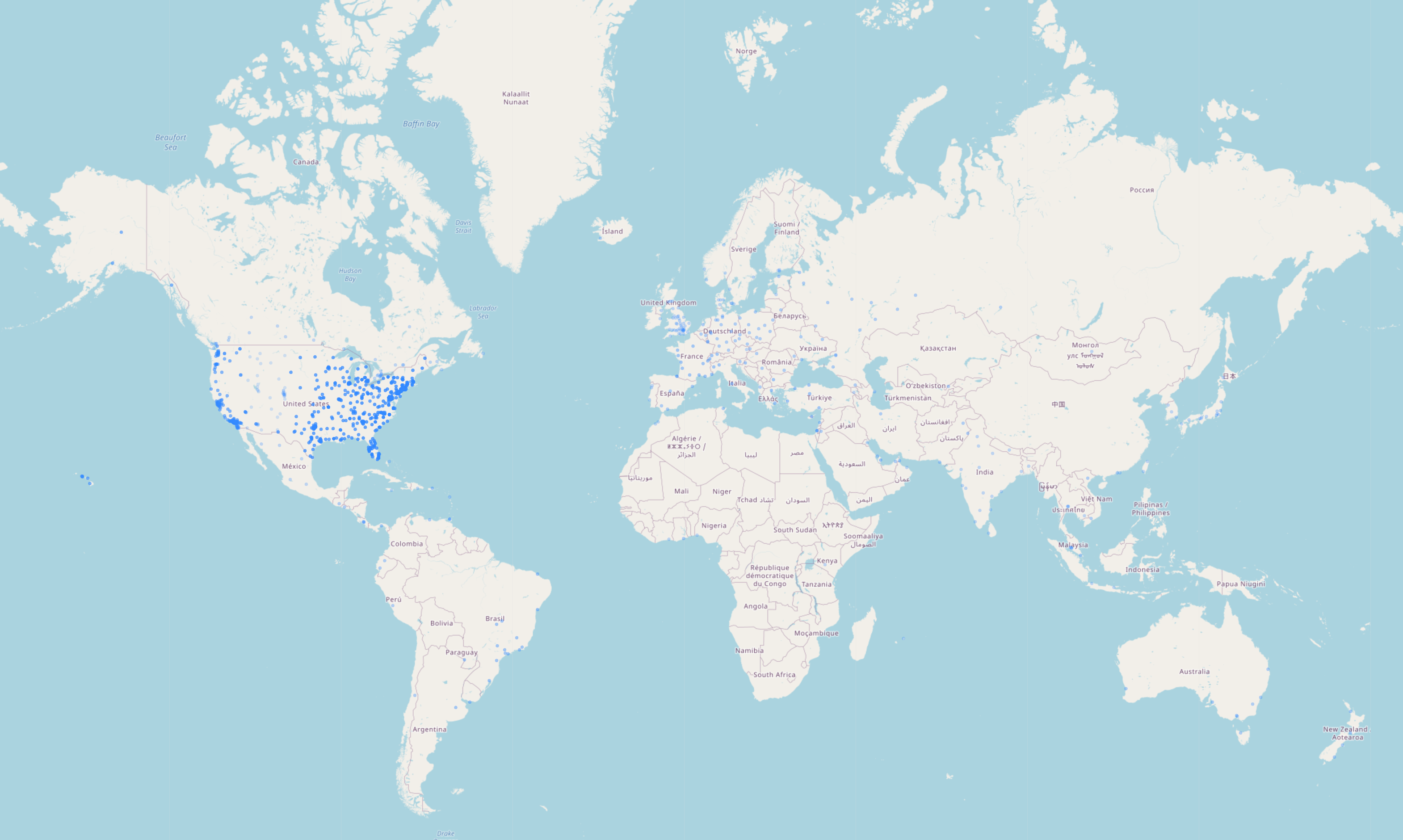}
        \caption{Fastly (IPv4)}
        \label{fig:sub3_v4}
    \end{subfigure}
    \begin{subfigure}{.33\textwidth}
        \centering
        \includegraphics[width=.9\linewidth]{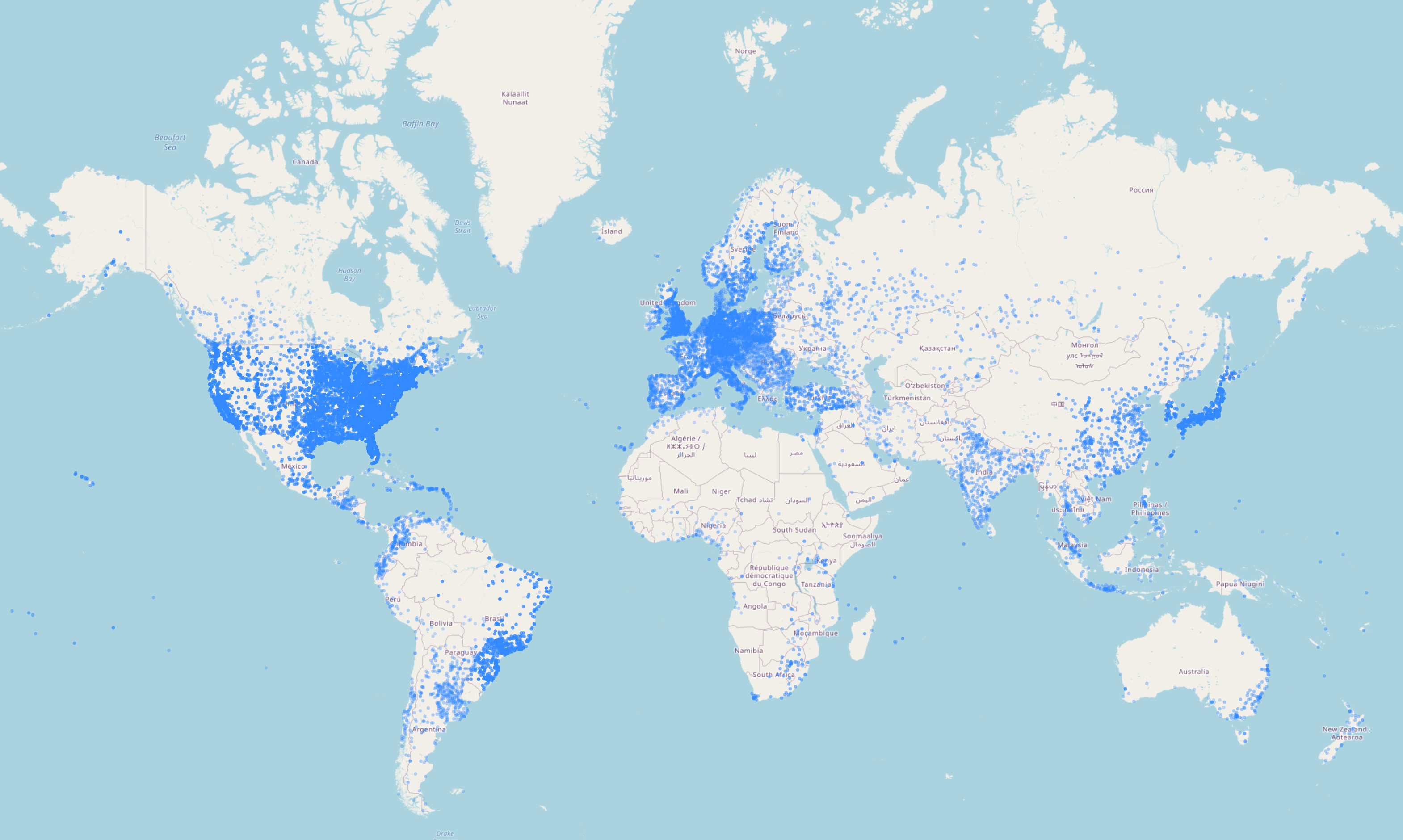}
        \caption{\akamaipr and \akamaiest (IPv6)}
        \label{fig:sub1_v6}
    \end{subfigure}
    \begin{subfigure}{.33\textwidth}
        \centering
        \includegraphics[width=.9\linewidth]{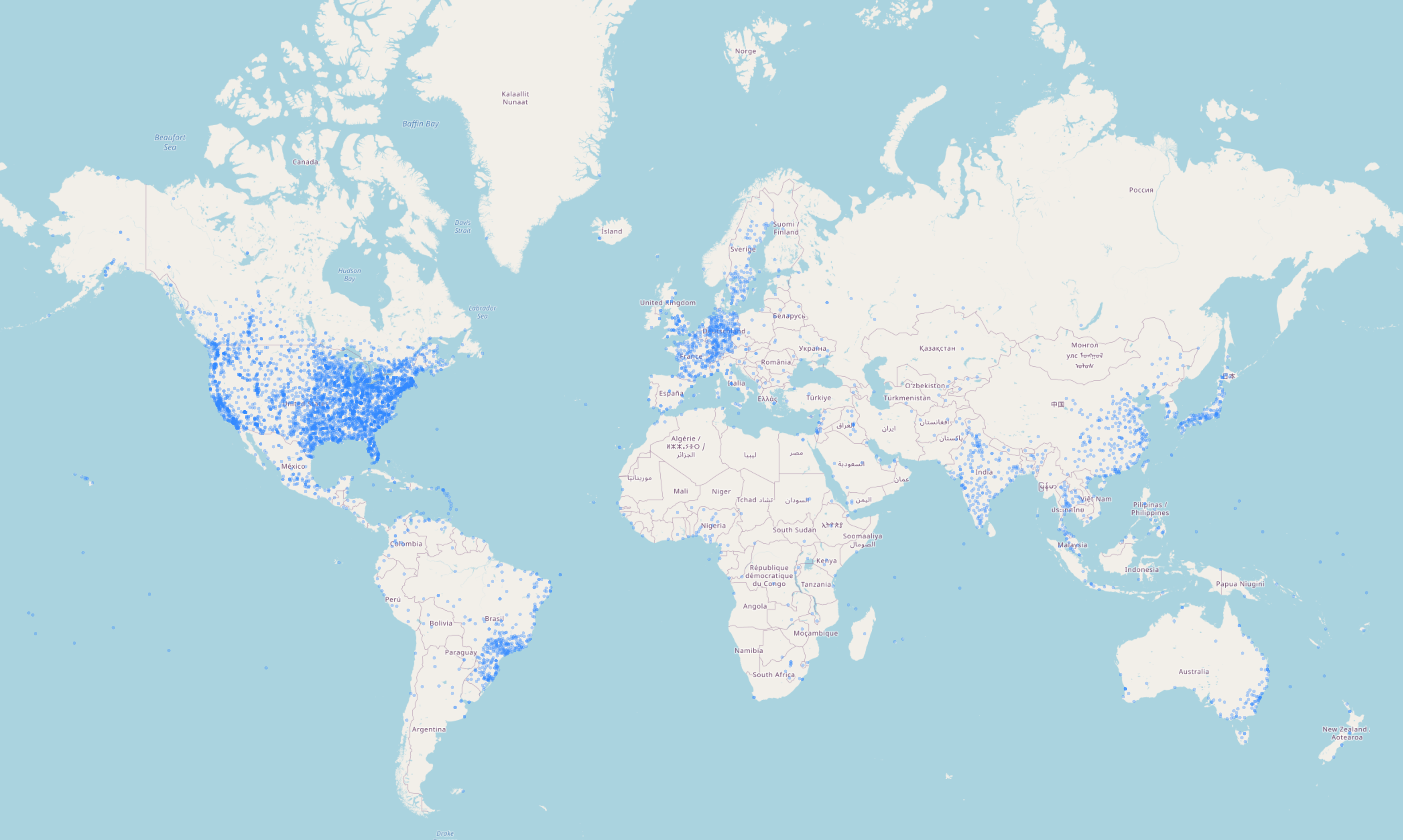}
        \caption{Cloudflare (IPv6)}
        \label{fig:sub2_v6}
    \end{subfigure}
    \begin{subfigure}{.33\textwidth}
        \centering
        \includegraphics[width=.9\linewidth]{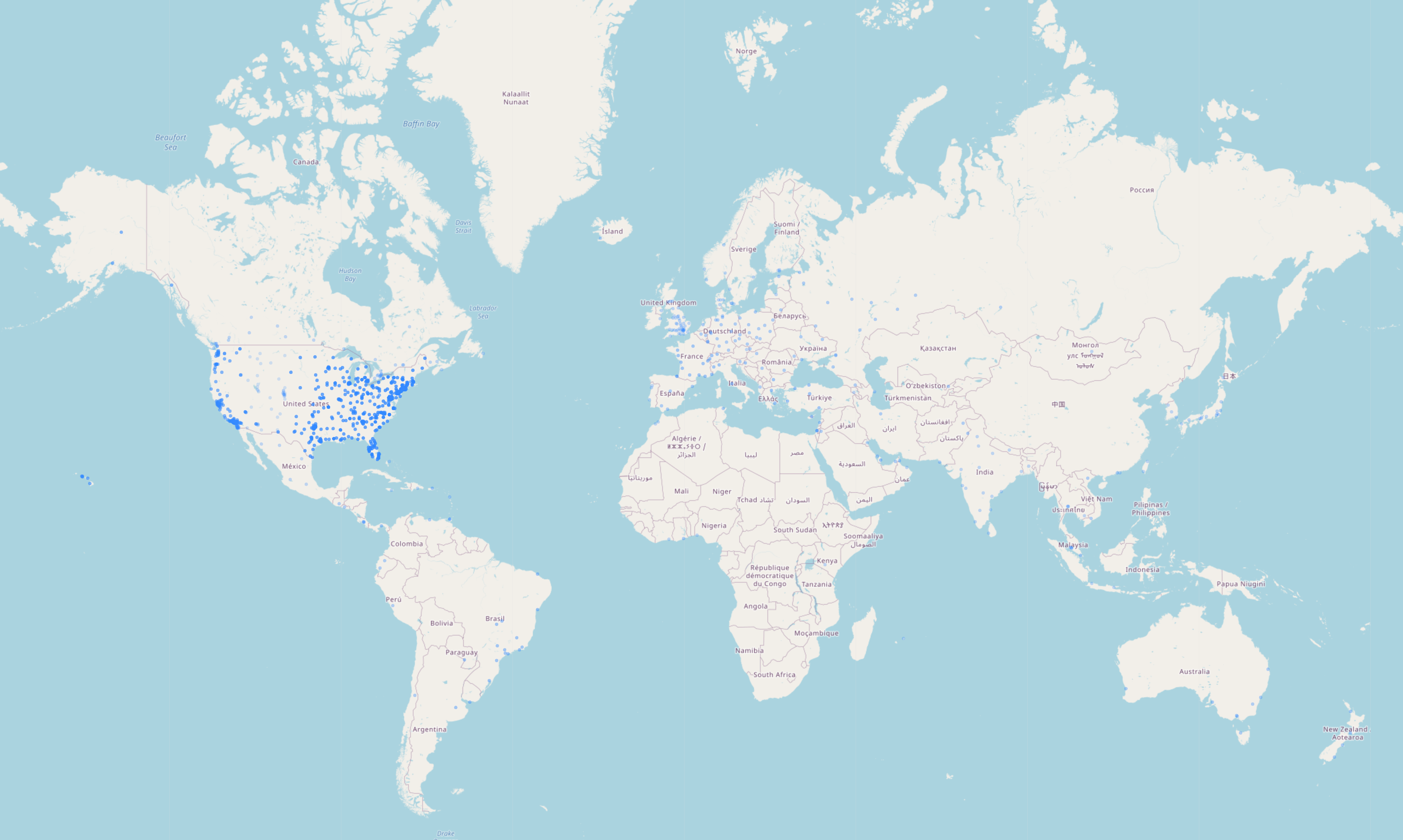}
        \caption{Fastly (IPv6)}
        \label{fig:sub3_v6}
    \end{subfigure}
    \caption{Geolocation of egress subnets per providing \ac{as} as published by Apple \cite{egress-nodes}\protect\footnotemark.}
    \label{fig:node_dist_ip_versions}
\end{figure*}

The important findings regarding egress node locations are collected in \Cref{sec:egress}.
The following presents numbers (\Cref{tbl:egress_asn_cities}) and visualizes these results in more detail, cities (\Cref{fig:citiesv4cdf,fig:citiesv6cdf}), covering \acp{cc} (\Cref{fig:ccv4cdf,fig:ccv6cdf}), and a visualization of egress node locations separated into IP versions (\Cref{fig:node_dist_ip_versions}).
The US are the top \ac{cc} which subnets got assigned to and a focus towards North America and Europe is currently visible.

\begin{table}[h]
    \setlength{\tabcolsep}{3.8pt}
\begin{threeparttable}
\footnotesize
\centering
\caption{Number of \acp{cc} by IPv4 subnets, IPv6 subnets and both combined.}
\label{tbl:egress_asn_cities}
\begin{tabular}{lrrr}
    \toprule
    & Covered Cities & Covered Cities IPv4 & Covered Cities IPv6 \\
    \midrule
    \akamaipr\tnote{1}      & \num{14088} & \num{853}  & \num{14085} \\
    \akamaiest\tnote{2}     & \num{7507}  & \num{455}  & \num{7507}  \\
    Cloudflare\tnote{3}     & \num{5228}  & \num{1134} & \num{5228}  \\
    Fastly\tnote{4}         & \num{848}   & \num{848}  & \num{848} \\
    \bottomrule
\end{tabular}
\begin{tablenotes}
\item[1] AS36183; \item[2] AS20940; \item[3] AS13335; \item[4] AS54113
\end{tablenotes}
\end{threeparttable}
\end{table}

\section{Additional Observations}
Apple introduced \ac{odoh} to describe \ac{dns} in \privaterelay.
\ac{dns} queries are sent encrypted through the first relay, similar to the HTTP requests but are then routed directly to the \ac{doh} server.
The client can learn its egress IP address and include it in the \ac{dns} queries \ac{ecs} information to receive an optimized response for the egress layer.

Given an active relay connection, the system ignores the local \ac{dns} resolver and uses its oblivious \ac{doh} server, \ie a \ac{doh} server connection through the relay system.
Currently, we identify Cloudlfare's public resolver~\cite{cloudflare-dns} as the one being used.

During our manual \privaterelay testing we observed that the service accepts the provided \ac{dns} records and connects to the corresponding ingress relay.
Nevertheless, after a short period of time we see that an additional QUIC connections is initiated.
In our observation, its target address is in the prefix (or \ac{as} in the dual stack case) of the configured ingress.
We assume these being backup or management connections to control the service on the client outside the actual service connection.

\footnotetext{Data by \copyright OpenStreetMap (http://openstreetmap.org/copyright), under ODbL (http://www.openstreetmap.org/copyright)}

\label{lastpage}
\end{document}